%% file: FLRW_with_GA.tex
\documentclass[11pt,a4paper]{article}

\usepackage{jcappub} 
\usepackage[utf8]{inputenc}
\usepackage[english]{babel}
\usepackage{amsmath,amsfonts,amssymb}
\usepackage{graphicx}
\usepackage{amsthm} 
\usepackage{empheq}
\usepackage{tensor}
\usepackage{hyperref}
\usepackage{subfig} 
\usepackage{xcolor}
\usepackage{cleveref}
\usepackage{bm} 
\usepackage[most]{tcolorbox}
\usepackage{booktabs}
\usepackage{afterpage} 

\definecolor{graylight}{cmyk}{.30,0,0,.67} 

\newcounter{examplecounter}

\theoremstyle{definition}

\theoremstyle{remark}

\numberwithin{equation}{section}

\theoremstyle{definition}
\theoremstyle{remark}

\newcommand{\T}[2]{\tensor{#1}{#2}}
\newcommand{\p}{\partial}
\newcommand{\Tn}[2]{\tensor*{#1}{#2}} 
\newcommand{\g}{\gamma}

\newcommand{\noi}{\noindent} 
\newcommand{\C}[1]{\mathcal{C}l_{#1}}

\newcommand{\dd}{\mathrm{d}}

\crefname{equation}{equation}{equations}
\crefname{equation}{Equation}{Equations}
\crefrangelabelformat{equation}{(#3#1#4--#5#2#6)}

\crefmultiformat{equation}{equations (#2#1#3}{, #2#1#3)}{#2#1#3}{#2#1#3}
\crefmultiformat{equation}{Equations (#2#1#3}{, #2#1#3)}{#2#1#3}{#2#1#3}

\title{Friedmann-Robertson-Walker spacetimes from the perspective of geometric algebra}

\author[a]{Pablo Bañón Pérez}
\author[a]{Maarten DeKieviet}
\author[b]{Bj{\"o}rn Malte Sch{\"a}fer}

\affiliation[a]{Physikalisches Institut, INF 226, Heidelberg University, 69120 Heidelberg, Germany}
\affiliation[b]{Astronomisches Recheninstitut, Zentrum f{\"u}r Astronomie der Universit{\"a}t Heidelberg, Philosophenweg 12, 69120 Heidelberg, Germany}

\emailAdd{pau.banon@physi.uni-heidelberg.de}

\abstract{The intention of our paper is to provide a pedagogical application of geometric algebra to a particularly well-investigated system: We formulate the geometric and dynamical properties of Friedmann-Robertson-Walker spacetimes within the language of geometric algebra and re-derive the Friedmann-equations as the central cosmological equations. Through the geometric algebra-variant of the Raychaudhuri equations, we comment on the evolution of spacetime volumes, before illustrating conformal flatness as a central property of Friedmann-cosmologies. An important aspect of spacetime symmetries are the associated conservation laws, for which we provide a geometric algebra formulation of the Lie-derivatives, of the Killing equation and of conserved quantities in Friedmann-Robertson-Walker spacetimes. Finally, we discuss the gravitational dynamics of scalar fields, with their particular relevance in cosmology, for cosmic inflation, and for dark energy.}

\begin{document}
	
\maketitle

\section{Introduction}
Geometric Algebra (GA) is an excellent tool for the mathematical description of many fields of physics \cite{Doran2013, Doran1994, Lasenby1996, Lasenby2009, Hestenes2003a_OerstedMedal, Hestenes_A_Unified_Language}, however, it has not yet been explored in detail in the realm of general relativity. In a previous work, we presented how the main mathematical tools of general relativity translate to GA \cite{perez_general_2024}: As tensor calculus and differential forms are naturally integrated into GA, the results are identical but with the benefit of simpler calculations and a clear geometric, and ultimately physical interpretation. Because technical calculations in relativity can be intimidating to students, looking for a new description providing a shorter, more concise and intuitive handling of the topic would be a great advantage. Note that the underlying theory is not modified at all, and given the fact that Friednmann-Robertson-Walker (FRW) spacetimes are incredibly well understood, it would be fair not to expect genuinely new results. Instead, we pursue the goal of formulating the dynamics of FRW-spacetimes in GA and juxtapose these results with the conventional Riemannian formulation of general relativity. In particular, the geometric properties of FRW-spacetimes as highly-symmetric and analytically solvable solutions to relativity become in our view very transparent in GA.

\Cref{sec:friedmanns-equations-from-the-metric} reiterates in GA the standard approach to FRW-spacetimes by using an intuitive comoving coordinate choice for formulating the metric and deriving Friedmann's equations. In the process, various elements and remarks on the difference between GA and tensor formalism or differential forms are presented, as well as geometric interpretations to the connection coefficients, the Riemann tensor and the Ricci tensor. Because the symmetries of these objects are naturally contained in the GA-formalism, the corresponding calculations are more straightforward compared to other formalisms.

Elaborating on the geometric nature of GA, \Cref{sec:friedmanns-equations-from-raychaudhuri-congruences} formulates the Raychaudhuri equations and uses them to derive Friedmann's second equation. Because the Raychaudhuri equations are naturally geometric, GA offers for each quantity a clear physical interpretation. Motivated by their important relationship to FRW-spacetimes, \Cref{sec:conformal-flatness-of-flrw-spacetimes} investigates conformal transformations in GA. Although the frame formalism provides some useful insights on conformal transformations, no particular computational advantage was found over Riemannian geometry with respect to showing the conformal invariance of the Weyl-tensor or the spatial flatness of FRW-spacetimes.

\Cref{sec:spacetime-symmetries-and-conservation-laws} is devoted to the symmetries and conservation laws of the FRW universe. Here, the tools needed to deal with Lie derivatives and Killing vectors are expressed in GA. The key result is that the necessity of a metric for the construction of Clifford spaces makes a proper, general definition of Lie derivatives a subtle problem, except for Killing vector fields. At the end, \cref{sec:quintessence-lagrangian-density}, we consider scalar field dynamics on FRW-backgrounds, as the central concept of cosmic inflation in the early Universe or the presence of dark energy in the late Universe. Quintessence models are easily expressed in GA and, in conjunction with Raychaudhuri equations, provide an alternative to deriving the slow-roll condition for an accelerated expansion to take place. 

We present a brief introduction to GA of flat spacetime in \cref{sec:introduction-to-ga}, and a thoroughly introduction to its use in GR in \cite{perez_general_2024}. In summary, GA offers a compact formalism to express GR. Its main benefit comes from its precision reflecting the intrinsic symmetries of the objects. This simplifies some calculations and permits a direct geometric interpretation of the results. The generalisation from the usual formulation of GR consists in the promotion of the tangent space $ T_p M $ from a vector space $ \mathcal{V} $ to a Clifford algebra, $ \C{1,3} $. Because the coordinate basis vectors are in general not orthonormal, in $ \C{1,3}$ we will define an orthonormal basis vector, known as tetrads. The tetrad basis represents the proper rest frame of an observer and allows for easy contractions of quantities.

In terms of notation, we will use $ \{g_\mu \} = \{\p_\mu x\} $ as coordinate vector basis, and $ \{\g_m\} $ as orthonormal tetrad basis frame. Correspondingly, the greek index  will refer to coordinates, $ \{\mu, \nu, \sigma, ...\} $, and Latin indices to tetrads $ \{m, n ,s, ...\} $. An object can have mixed indices, e.g.: $ R_{\mu \nu m n} $. In the case of having to write an explicit index, the hat will denote the tetrad components $ \{\hat{t},\hat{ r},\hat{\theta}, \hat{\phi}\} $.

\section{Friedmann's equations with coordinate intuition}\label{sec:friedmanns-equations-from-the-metric}

\subsection{Choice of frame}
We start with the FRW metric expressed in hyperspherical (or curvature-normalised) coordinates obtained under the assumptions of homogeneity and isotropy of the Universe,
\begin{equation} \label{eq:FRW_Metric}
	\dd s^2 = \dd t^2 -a^2 (t) \left[ \frac{\dd r^2}{\sqrt{1 - k r^2}} + r^2\dd\Omega^2\right].
\end{equation}
Our hyperspherical coordinates are $ (t,  r, \theta, \phi) $, and $ k $ is the usual constant quantifying the curvature of the universe, with the three cases $k = -1$ (open), $k=0$ (flat) and $k = -1$ (closed).

Suggested by the line element \Cref{eq:FRW_Metric}, the usual choice of basis for the tangent space $T_p M$ is the coordinate basis $ \{g_\mu\} $. This choice of basis has the problem of not being orthonormal, as it can be seen by projecting them into each other,
\begin{equation}
	g_\mu \cdot g_\nu = g_{\mu\nu}.
\end{equation}

It is known in the literature that for the representation of the basis of a proper observer, we have to change to an orthonormal frame, called a tetrad. We will denote the tetrad frame by  $ \{ \g_m \} $. This choice is made to highlight the Minkowskian geometry of the tangent space $T_pM$. The usage of an orthonormal frame simplifies contractions and provides a clear distinction in the dependence of objects on coordinates or on vector basis, which remains a problem frequently encountered in GR literature. 

The operator relating the coordinate and tetrad basis of $ T_p M $ is called the vierbein. We express it in matrix form as $ \T{e}{^m_\mu} $, so the relationship between frames is
\begin{equation}\label{eq:Tetrad_Vierbein}
	g_\mu = \T{e}{^m_\mu} \g_m,
\end{equation}
allowing us to write the line element as
\begin{equation}
\dd s^2  = \eta_{ m n} \T{e}{^m_\mu}  \T{e}{^n_\nu} \dd x^\mu \dd x^\nu,
\end{equation}
where $\eta_{ m n} = \text{diag}(+1, -1, -1, -1 )$ is the Minkowski metric.

\begin{figure}
	\centering
	\def\svgwidth{0.4\textwidth}
	\input{Basis_transformation.pdf_tex}
	\caption{Picture illustrating the tetrad formalism. When defining the basis of the tangent space $T_p M$ at point $ p $, we have the freedom to choose the basis we want to work on. The \textit{natural} basis would be the coordinate basis $g_\mu = \p_\mu x$, but we can also form an orthonormal basis $\{\g_m\}$ called a tetrad. The two basis are related by the vierbein $\T{e}{^m_\mu}$.} 
	\label{fig:Basis_Transformation}
\end{figure}
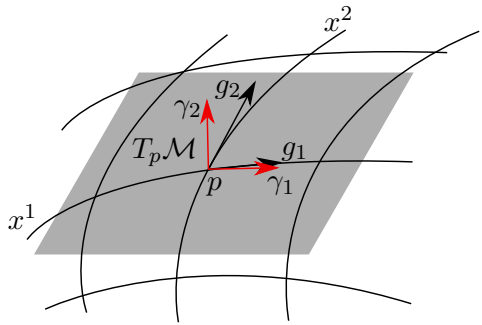

Colloquially speaking, we can think about the vierbein as the \textit{square root of the metric}. In the FRW case, the coordinate basis vectors are already orthogonal, so we simply need to normalize them to obtain the tetrad base. The matrix form of the vierbein for our FRW spacetime is
\begin{equation}
	[\T{e}{^m_ \mu}] = 
	\begin{pmatrix}
		1 & 0 & 0 & 0 \\
		0 & \frac{a}{\sqrt{1-k r^2}} & 0 & 0 \\
		0 & 0 & a r & 0 \\
		0 & 0 & 0 & a r \sin\theta
	\end{pmatrix}.
\end{equation}
From the conditions $ \T{e}{^m_\nu} \T{e}{_m^\mu} = \delta^\mu_\nu $ and $ \T{e}{^m_ \mu} \T{e}{_n^\mu} = \delta^m_n $, we can obtain the inverse vierbein
\begin{equation}
	[\T{e}{_m^\mu}] = 
	\begin{pmatrix}
		1 & 0 & 0 & 0 \\
		0 & \frac{\sqrt{1- k r^2}}{a} & 0 & 0 \\
		0 & 0 & (a r)^{-1} & 0 \\
		0 & 0 & 0 & (a r \sin\theta)^{-1}
	\end{pmatrix}.
\end{equation}

All this considered, the tetrad frame $ \{\g^m\} $ can be obtained in terms of the coordinate base as
\begin{equation}\label{eq:Coord_Tetrad}
	\begin{rcases}
		g_\mu \cdot g_\nu = g_{\mu\nu}\\
		g^\mu g_\nu = \delta_\nu^\mu\\
		\g_m \cdot \g_n = \eta_{mn}\\
		\g^m \g_n = \delta_n^m\\
		g_\mu = \T{e}{^m_ \mu}\g_m
	\end{rcases}
	\quad	\Rightarrow \quad 
	\begin{rcases}
		\g^t = g^t\\
		\g^ r = \frac{a}{\sqrt{1-k r^2}} g^ r\\
		\g^\theta = a r g^\theta\\
		\g^\phi = a r \sin\theta g^\phi
	\end{rcases}
\end{equation}

The tetrad basis vectors  $ \{ \g_m \} $, and their reciprocals,  $ \{ \g^m \} $, are related by the Minkowski metric $\eta_{mn} = \text{diag}$(+1, -1, -1, -1), in our choice of signature, $\g_m = \eta_{mn} \g^n$.

\subsection{Bivector connection coefficients}
The next element needed to obtain Friedman's equations are the connection coefficients. Because tetrad frames are orthonormal, the relationship between the tetrad at a point $p$ and the parallel-transported tetrad is necessarily a Lorentz transformation. Therefore, the connection coefficients in the tetrad base are the generators of Lorentz transformations, which in GA are bivectors, 

In particular, we can interpret the connection coefficients $\omega$ as a linear map from vectors to bivectors
\begin{equation}
	\omega: \Lambda^1(\mathcal{V})  ~\rightarrow~ \Lambda^2(\mathcal{V}), ~ v \in \Lambda^1(\mathcal{V}) \mapsto \omega(v) \in \Lambda^2(\mathcal{V}),
\end{equation}
where $\Lambda^i(\mathcal{V})$ is the exterior algebra of grade $i$. The geometric meaning of the connection coefficients provided by GA is the following: Considering $ v $ a particular direction of the space, the bivector $ \omega(v) $ provides the rotation (or boost) of the frame when parallel transported in the $ v $ direction. That's the reason one refers to them as \textit{rotation coefficients}.

Their action over a multivector $M$ can be obtained by performing an infinitesimal Lorentz transformation in the direction $g_\mu$. The bivector generating the corresponding transformation will be $\omega(g_\mu) = \omega_\mu$. And the corresponding rotor will be $ R = \exp(\epsilon \omega_\mu /2) \approx 1 + \frac{1}{2} \epsilon \omega_\mu $, when considering a parameter $ \epsilon\ll 1 $. To perform a rotation in GA one needs to sandwich the multivector with the corresponding rotor. In the case of an infinitesimal transformation, this double-sided action can be written as
\begin{equation}
	R M \tilde{R} \approx M + \epsilon \frac{1}{2}\left[\omega_\mu, M \right],
\end{equation}
where $ \left[\omega_\mu, M \right] = \omega_\mu M - M \omega_\mu$ is the commutator. Therefore, the action of $ \omega_\mu $ on any multivector must take the form of a commutator, and we can write the covariant directional derivative of any multivector as
\begin{equation}\label{eq:Cov_Deriv_Multivector}
	D_\mu M = \p_\mu M + \frac{1}{2}\left[\omega_\mu, M\right].
\end{equation}

In analogy to the vector derivative operator $\nabla = \g^\mu \p_\mu$, utilizing the covariant directional derivative we can define the covariant vector operator 
\begin{equation}
	D = g^\mu D_\mu.
\end{equation}
There are two worthwhile things to notice from $D$. First, it is a coordinate-free object that can be decomposed in any system of coordinates, second, algebraically we can treat it as a vector and use the inner, wedge and geometric products with it \cite{perez_general_2024}.

The directional covariant derivative is obtained by projecting $D$ into the desired direction $a$
\begin{equation}
	a \cdot D = a^\mu D_\mu.
\end{equation}

The commutator of a bivector with a vector is their scalar product. The action of the connection coefficients over the tetrad basis is 
\begin{equation}\label{eq:Connection_tetrad}
	D_\mu \g_n = \frac{1}{2}[\omega_\mu, \g_n] = \omega_\mu \cdot \g_n.
\end{equation}

We can directly solve \Cref{eq:Connection_tetrad} for the torsion-free case and calculate the connection coefficients in the coordinate direction $g_\mu$ as a function of the metric
\begin{equation}\label{eq:Connection-coefficients}
	\omega_{\mu}=\omega\left( g_{\mu}\right)=\frac{1}{2}\left[g^{\nu} \wedge D\left(g_{\mu\nu}\right)+g^{\sigma} \wedge \partial_{\mu} g_{\sigma}\right].	
\end{equation}

In case of a diagonal metric, meaning orthogonal coordinates, and by choosing a tetrad frame aligned with the coordinate frame, 
\begin{equation}
	\g_\mu = \frac{g_\mu}{|g_{\mu\mu}|^{1/2}},
\end{equation}
the last term of \cref{eq:Connection-coefficients} vanishes, producing a very simple expression to obtain the connection bivectors from the metric
\begin{equation}\label{eq:Connection-coefficients-diagonal}
	\omega_{\mu}=\omega\left( g_{\mu}\right)=\frac{1}{2}g^{\mu} \wedge g^\lambda \p_\lambda g_{\mu\mu}.	
\end{equation}

\Cref{eq:Connection-coefficients-diagonal} allows us to obtain directly the connection coefficients from the metric. In comparison with the Christoffel formula, this expression is easier to apply, and, in comparison with the \textit{guess-and-check} method of differential forms \cite{Misner1973}, it is general and straightforward.

If we expand $ \omega_\mu $ in the tetrad basis, we obtain the so-called Ricci rotation coefficients.

\begin{equation}	\label{eq:Bivector_Rot_coef}
	\omega_\mu = \frac{1}{2} \omega_{mn\mu} \g^m \wedge \g^n.
\end{equation}

Notice that because $\omega_\mu $ is a bivector, its components are automatically anti-symmetric in their first two indices $ \omega_{m n \mu} = -\omega_{n m \mu}$. This is an expected symmetry for the generator of a Lorentz transformation.

This decoupling of the degrees of freedom has important consequences on the number of connection coefficients at our disposal. Whereas the Christoffel symbols have $40$ degrees of freedom in a torsion-free space, here, because the connection coefficients map vectors to bivectors there can only be $ 4 \times 6 = 24 $ of them, corresponding to $ 3 $ rotations with generators $\g_i \wedge \g_j $, and $3$ boosts with generators $\g_0 \wedge \g_i$, in each of the $ 4 $ possible directions of displacement.

The remaining $ 16 $ degrees of freedom needed to cover all $ 40 $ of the Christoffel symbols are encoded in the vierbein $ \T{e}{_m^\mu} $, and they originate in the freedom of choice of coordinates and the tetrad frame. Thus, with the tetrad description of GR, we can decouple the degrees of freedom related to coordinate choice and the ones related to frame choice.

Finally, we should note that because the Chrystoffel symbols are not tensors, they are not related to the components $\omega_{m n \mu}$ by a change of base usage of the vierbein $\T{e}{_m^\mu}$. The relationship between them is more complicated and includes derivatives of the vierbein.

We can now calculate the connection coefficients for the FRW universe using \Cref{eq:Connection-coefficients-diagonal} and obtain
\begin{equation}
		\begin{aligned}
	\omega_{t} & = \frac{1}{2}g^{t} \wedge g^\lambda \p_\lambda g_{tt}= 0\\
	\omega_{ r} & = \frac{1}{2}g^{ r} \wedge g^\lambda \p_\lambda g_{ r  r} = \frac{ \dot{a}}{\sqrt{1-k r^2}} \g^t \wedge \g^r\\
	\omega_{\theta} & =\frac{1}{2} g^{\theta} \wedge g^\lambda \p_\lambda g_ {\theta \theta} =  \sqrt{1-k r^2}\g^r \wedge \g^\theta + \dot{a} r \g^r \wedge \g^\theta\\
	\omega_{\phi} & =\frac{1}{2} g^{\phi} \wedge  g^\lambda \p_\lambda  g_{\phi\phi}\\
	 & = r \sin (\theta ) \dot{a} \g^t \wedge \g^\phi +  \sin (\theta ) \sqrt{1-k r^2} \g^r \wedge \g^\phi +  \cos (\theta ) \g^\theta \wedge \g^\phi.
	\end{aligned}
\end{equation}
As a reminder, overdot represents a derivative with respect coordinate time $\dot{a} = \p_t a$.

In an FRW universe in GA, we have 4 bivectors, with a total of $ 6 $ coefficients. A noticeable reduction in comparison with the $ 13 $ Christoffel symbols of tensor calculus.

The geometric interpretation of the connection coefficients is can be observed by analyzing the parallel-transport equation of a vector $v$ in the $a$ direction
\begin{equation}
		a \cdot D v = (a^\mu D_\mu) (v^m \g_m) = 
		a^\mu (\p_\mu v^m + v^n \T{\omega}{^m_{n \mu}})\g_m = 0 \Rightarrow 
		a^\mu \p_\mu v^m = - a^\mu v^n \T{\omega}{^m_{n \mu}}.
\end{equation}
A vector $v$ is parallel-transported in the $a$-direction if the variation in the vector components is precisely compensated by the transformation of the transported frame. In the case of tetrads, the transformation of the frame is a Lorentz transformation, so it is not necessary to take into account any possible variation in the lengths of the basis vectors due to coordinates.

\begin{figure}
	\centering
	\def\svgwidth{0.6\textwidth}
	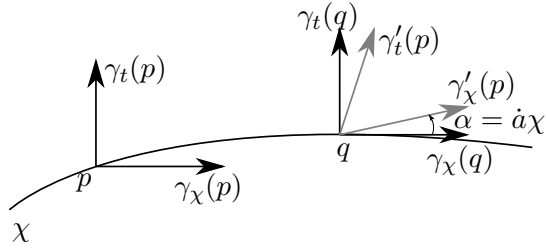
	\caption{Picture to illustrate the transformation of a frame when parallel-displaced in the $ r$ direction. At point $p$ we have an inertial frame $\{\g_\mu(p)\}$. The relationship between the local frame at $q$, $\{\g_\mu(q)\}$ and the parallel-displaced frame $\{\g'_\mu(p)\}$ in grey, is a Lorentz transformation with rapidity $\dot{a} $. This is given by the connection coefficient $\omega_ r = \dot{a}\gamma^{t} \wedge \gamma^{ r}$} 
	\label{fig:Example_Connection}
\end{figure}

In the FRW case, and taking the coefficient $\omega_ r =\dot{a} \gamma^{t} \wedge \gamma^{ r} $ as an example we can read that if we parallel-transport an orthonormal frame along the $ r$ coordinate, it will experience a boost with rapidity $\dot{a} r$, see \Cref{fig:Example_Connection}. Notice that $\omega_t = 0$, meaning that frames parallel-displaced in the $t$ direction will suffer no rotation or boost.

\subsection{Riemann curvature}
In a torsion-free space, we can express the commutator of covariant derivatives by the operator $D \wedge D$
\begin{equation}
	D \wedge D = g^\nu \wedge g^\mu [D_\nu, D_\mu].
\end{equation}

Applying $D \wedge D$ to a multivector $M$ and expanding into components using \Cref{eq:Cov_Deriv_Multivector}, we obtain the action of the Riemann tensor in GA
\begin{equation}\label{eq:Cocurl_Multivector}
	D \wedge D M = g^\mu \wedge g^\nu [D_\mu, D_\nu] M = g^\mu \wedge g^\mu [\mathbf{R}(g_\mu \wedge g_\nu), M].
\end{equation}
Where we have defined the Riemann tensor $\mathbf{R}(g_\mu \wedge g_\nu)$ from the connection coefficients as
\begin{equation}\label{eq:Riemann_components}
	\mathbf{R}(g_\mu \wedge g_\nu) = \mathbf{R}_{\mu\nu} = \p_\mu \omega_\nu - \p_\nu \omega_\mu + \frac{1}{2} [\omega_\mu, \omega_\nu].
\end{equation}

The Riemann tensor in GA is a map from bivectors to bivectors, 
\begin{equation}
	\mathbf{R}: \Lambda^2(\mathcal{V}) \rightarrow \Lambda^2(\mathcal{V}), ~ B \in \Lambda^2(\mathcal{V}) \mapsto \mathbf{R}(B) \in \Lambda^2(\mathcal{V}).
\end{equation}

\begin{figure}
	\centering
	\def\svgwidth{0.7\textwidth}
	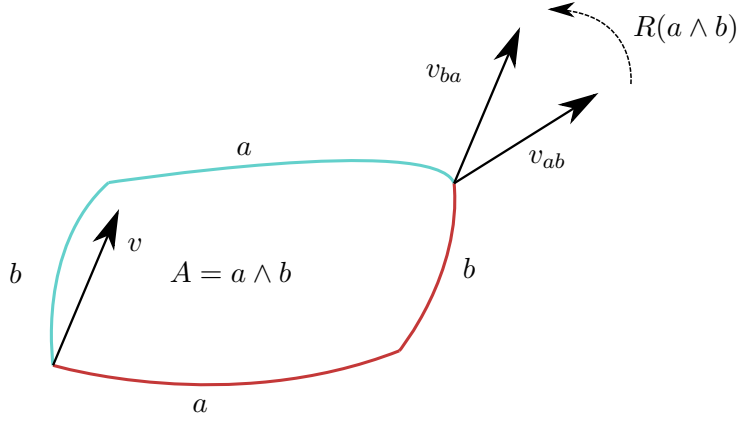
	\caption{Representation of the effect of transporting vector $ v $ to the same point through two different paths. When transported through the red path, $ a \rightarrow b $, the resulting vector is $ v_{ab} $. When transported through the blue path, $ b \rightarrow a $, the resulting vector is $ v_{ba} $. $ v_{ab} $ and $ v_{ba} $ are related by a rotation that is a function of the area spanned between the paths, $ A = a \wedge b $. That function is called the Riemann tensor.} 
	\label{fig:Curvature}
\end{figure}

Considering that bivectors represent both areas and are the generators of rotations, the geometrical meaning of the Riemann tensor is very clear: It relates a differential area with the rotation experienced by a vector when parallel-displaced along its contour, as illustrated by \Cref{fig:Curvature}.

We can expand the Riemann bivector in the coordinate basis or in the tetrad basis. Obtaining the usual tensorial expression of the Riemann $\mathbf{R}_{\mu \nu \alpha \beta}$ or a mixed indices expression  $\mathbf{R}_{\mu \nu m n}$.
 \begin{equation}
	\mathbf{R} (g_\mu \wedge g_\nu) = \frac{1}{2} \mathbf{R}_{\mu \nu \alpha \beta} (g^\alpha \wedge g^\beta) = \frac{1}{2} \mathbf{R}_{\mu \nu m n} (\g^m \wedge \g^n)
\end{equation}

What is remarkable is that, when expressed in this manner, the symmetries of the Riemann become evident, and the interpretation of its two pairs of different indices is obtained: Two of them are related to the coordinate area, and the other two are related to the rotation in the tangent space.

When formulated in GA, the Riemann tensor gives right away its symmetries and the correct number of degrees of freedom.

\begin{enumerate}
	
	\item The first two components of $ \T{\mathbf{R}}{_{\mu\nu m n}} $ are related to coordinates and must be anti-symmetric.
	
	\item The same applies to the second pair of indices, they must be anti-symmetric because they correspond to bivector indices.
	
	\item Because it is a function mapping bivectors to bivectors, in a space of 4 dimensions it can have at most $ 6 \times 6 = 36 $ degrees of freedom. However, because the Riemann tensor is an injective function, we can interchange the two pairs of indices, reducing the degrees of freedom further $ 36 - (30/2) = 21 $.
	
	\item The remaining $ 21 $ degrees of freedom can be reduced further if we take into account the cyclic permutation of the last three indices. It only holds in spaces without torsion and is a consequence of the protractionlessness of the Riemann tensor \cite{Hestenes_STG_with_GC} which in GA can be expressed as
	\begin{equation}
		\p_a \wedge \mathbf{R}(a \wedge b) = 0
	\end{equation}
	or, expanded in the form of first algebraic Bianchi identity,
	\begin{equation}
		\mathbf{R}(a \wedge b) \cdot c + \mathbf{R}(c \wedge a) \cdot b + \mathbf{R}(b \wedge c) \cdot a = 0,
	\end{equation}
	
	\noi producing a total of $ 21 - 1 = 20 $ degrees of freedom.
	
\end{enumerate}

The immediacy in the acquisition of the symmetries of the Riemann in GA should be compared with the amount of calculation required to obtain the same results in the tensorial formulation.

In the case of the FRW universe, the components of the Riemann tensor take a particularly compact form in the tetrad basis. Which we obtain with $\mathbf{R}_{mn} = \T{e}{_m^\mu} \T{e}{_n^\nu} \mathbf{R}_{\mu \nu}$
\begin{equation}
	\begin{aligned}
		&\mathbf{R}_{\hat{t} m}= \frac{\ddot{a}}{a} \gamma^{\hat{t}} \wedge \gamma^{m}, \quad m=\hat{ r}, \hat{\theta}, \hat{\phi} \\
		&\mathbf{R}_{mn}=-\frac{\left(\dot{a}^2+k\right)}{a^2} \g^m \wedge \g^n, \quad   m, n=\hat{ r}, \hat{\theta}, \hat{\phi}
	\end{aligned}
\end{equation}

\subsection{Ricci vector, Ricci scalar and Einstein tensor}
The Ricci tensor appears naturally from \Cref{eq:Cocurl_Multivector} when acting on a vector, $M \rightarrow a$,
\begin{equation}
	D \wedge D a = g^\mu \wedge g^\mu [\mathbf{R}(g_\mu \wedge g_\nu), a] = g^\mu \wedge g^\mu \mathbf{R}(g_\mu \wedge g_\nu)\cdot a = R(a),
\end{equation}
where $R(a) = \T{R}{_{\beta \mu}} a^\beta g^\mu$. In GA, the Ricci tensor is a map from vectors to vectors
\begin{equation}
	R: \Lambda^1(\mathcal{V}) \rightarrow \Lambda^1(\mathcal{V}), ~ v \in \Lambda^1(\mathcal{V}) \mapsto R(v) \in \Lambda^1(\mathcal{V}).
\end{equation}

For any direction $a$ it returns a vector, $ R(a) $, the dual of which is a 3-volume, $ V = R(a)I $. The Ricci vector quantifies the variation of $V$ due to curvature when displaced in the $ a $-direction

That is, in a similar manner that the Riemann tensor governs the evolution of a vector or a displacement parallel propagated along a geodesic, the Ricci tensor governs the corresponding evolution of a small volume \cite{Loveridge2004}: If $ a = \dd x/\dd \tau$ and $V = R(a)I$, then
\begin{equation}
	\frac{D^2}{\dd \tau^2} \delta V - \frac{D_{flat}^2}{\dd\tau^2} \delta V = - \delta V R(v ) \cdot v
\end{equation}

To calculate the components of the Ricci vector, we have to contract the Riemann tensor. This task can be performed easily in the tetrad frame due to its orthonormality
\begin{equation}
	\begin{aligned}
		R_{\hat{t}} & = \g^n \cdot \mathbf{R}_{n \hat{t}} = - 3 \frac{\ddot{a}}{a} \g^t\\
		R_{m} & = \g^n \cdot \mathbf{R}_{n m} = \frac{\left(a\ddot{a} + 2 \left(\dot{a}^2+k\right)\right)}{a^2} \g^m, \quad   m=\hat{ r}, \hat{\theta}, \hat{\phi}
	\end{aligned}
\end{equation}

To obtain the Ricci scalar, we can perform either a second contraction with the Ricci vector or a direct bivector contraction with the Riemann. Just as before, contraction in the tetrad frame is easier
\begin{equation}\label{eq:Ricci_Scalar_FLRW}
	\begin{aligned}
		\mathcal{R} & =\left(\gamma^{n} \wedge \gamma^{m}\right) \cdot \mathbf{R}_{m n} = 
		- 6 \frac{\left(a \ddot{a}+\dot{a}^2+k\right)}{a^2}
	\end{aligned}
\end{equation}
and recovers the correct result obtained from tensor calculus. Notice that, because the products in GA are non-commutative in general, one should be careful and perform the contractions in the right order, $ \left(\gamma^{n} \wedge \gamma^{m}\right) \cdot \mathbf{R}_{m n} = - \left(\gamma^{m} \wedge \gamma^{n}\right) \cdot \mathbf{R}_{m n} $.

We are now in disposition to calculate the Einstein vector, which in GA is a map from vectors to vectors 
\begin{equation}
	G: \Lambda^1(\mathcal{V}) \rightarrow \Lambda^1(\mathcal{V}), ~ v \in \Lambda^1(\mathcal{V}) \mapsto G(v) \in \Lambda^1(\mathcal{V}).
\end{equation}
The interpretation of the Einstein vector is that given a direction $a$, $G(a) \cdot a$ is the curvature of the 3-space perpendicular to $a$ \cite{Loveridge2004}. The Einstein vector is obtained by the usual combination
\begin{equation}\label{eq:Einstein_Tensor}
	G(a) = R(a) - \frac{1}{2} a \mathcal{R}.
\end{equation}
When calculated for the FRW case one obtains
\begin{equation}
	\begin{aligned}
		& G_{\hat{t}} = 3\frac{ (k + \dot{a}^2)}{a^{2}} \g^{t}\\
		& G_{m} = - \frac{\left(2 a \ddot{a} + \dot{a}^2+k\right)}{a^2} \g^m, \quad m =  \hat{r}, \hat{\theta}, \hat{\phi}
	\end{aligned}
\end{equation}
For the FRW case, this means that the 3-space has curvature given by
\begin{equation}
	G(g_t) \cdot g_t = G(\T{e}{^{\hat{t}}_t} \g_t)\cdot (\T{e}{^{\hat{t}}_t} \g_t) = G_{\hat{t}} \cdot \g_t = 3 \frac{(k + \dot{a}^2)}{a^{2}}.
\end{equation}
The fact that $G(g_t) \cdot g_t$ only depends on $t$, reflects the fact that, for a given time slice, the curvature of the 3-space is constant in all directions.

\subsection{Gravitational field equations}
In GA, Einstein's equations are
\begin{equation} \label{eq:Einstein_Eq}
	G(a) - \Lambda a = 8 \pi T(a).
\end{equation}

For a perfect fluid, the expression for the energy-momentum tensor is \cite{Doran2013}
\begin{equation}\label{eq:EM_Tensor_Perfect_Fluid}
	T(a) = (\rho + p) a \cdot u u - p a.
\end{equation}
Where $ \rho $ is the energy density, $ p $ the pressure, and $ u $ is the 4-velocity of the fluid. We can make a particular choice of a frame where the fluid is at rest and therefore $u = \g^t $. From the time component of Einstein's equations, we get the first of Friedmann's equations
\begin{equation}\label{eq:Friedmann1}
	\gamma_{t}\left(3\left(\frac{\dot{a}}{a}\right)^{2}+3 \frac{k}{a^{2}}\right)-\Lambda \gamma_{t}=8 \pi \rho \gamma^{t} \Rightarrow \left(\frac{\dot{a}}{a}\right)^{2}=\frac{8 \pi}{3} \rho-\frac{k}{a^{2}}+\frac{\Lambda}{3}
\end{equation}

Calculating now any spatial component of Einstein's equations, we get the second of Friedmann's equation
\begin{equation}\label{eq:Friedmann2}
	\begin{split}
		\left(2 \frac{\ddot{a}}{a}+a^{-2}\left(k+\dot{a}^{2}\right)\right) \gamma_ r-\Lambda \gamma_{ r}=-8 \pi p \gamma_{ r}\\
		2 \frac{\ddot{a}}{a}= -\left(\frac{\dot{a}}{a}\right)^{2}-8 \pi p - \frac{k}{a^{2}}+\Lambda.
	\end{split}
\end{equation}

\section{Friedmann's equations from Raychaudhuri congruences}
\label{sec:friedmanns-equations-from-raychaudhuri-congruences}
An alternative and geometrically intuitive way of obtaining Friedmann's equations is from Raychaudhuri's equation. It describes the evolution of a congruence of geodesics by quantifying the evolution of their enclosed volume. We would like to point out that the Raychaudhuri-equation is simply a statement about the motion of a bundle of test particles through spacetime. How they will actually evolve and be accelerated relative to each other will be determined by the geometrical properties of the spacetime, which in turn are determined by the field equations.

In the FRW-case, the Raychaudhuri congruence has an intersection point as $a \to 0$ in the finite past, and given the current values of the densities and equations of state of the cosmological fluids, no further intersection point in the future. In an FRW-universe, shear and vorticity vanish due to the cosmological symmetries, as these terms effectively introduce anisotropies. Writing Raychaudhuri's equation as the evolution of the volume $\theta = \nabla_\mu u^\mu$ with proper time $\tau$ leads to:
\begin{equation}\label{eq:Raychaudhuri_Eq}
	\frac{\dd \theta}{\dd \tau} = 
	- \frac{\theta^2}{3} - R_{\mu\nu}u^\mu u^\nu + \nabla_\mu (u^\nu \nabla_\nu u^\mu),
\end{equation}
where in FRW-spacetimes  proper time, as the length of geodesic world lines, is equal to the coordinate or cosmic time $t$.

In Raychaudhuri's equation, the decomposition of the Riemann curvature into the Ricci- and Weyl-tensors has a particular intuitive interpretation, as only the Ricci curvature can change the enclosed volume. Weyl-curvature, which is absent in FRW spacetimes, would be responsible for a change in the shape of the enclosed volume, in violation of the cosmological symmetries. The Hubble expansion causes volumes to change proportionally to $a^3(\tau) = a^3(t)$, where $a(t)$ as a function depends on the densities and the equations of states of the cosmological fluids as encapsulated by the energy-momentum tensor, and by virtue of the field equation, by the Einstein-tensor, which reflects only Ricci-curvature.

\subsection{Raychaudhuri congruences}\label{sec:Raychaudhuri-congruences}
The direct reformulation of \Cref{eq:Raychaudhuri_Eq} in terms of GA is
\begin{equation}\label{eq:GA_Raychaudhuri_Eq_1}
	\frac{\dd \theta}{\dd \tau} = - \frac{\theta^2}{3} - R(g_t) \cdot g_t+ D \cdot \left(u \cdot D u\right),
\end{equation}
with $ u $ being a vector tangent to a geodesic. Their divergence $\theta $ is given by
\begin{equation}\label{eq:Divergence}
	\theta  = D \cdot u = D_m u^m = D_\mu u^\mu = \frac{1}{\sqrt{|g|}}
	\frac{\partial\left(\sqrt{|g|} u^{\mu}\right)}{\partial x^{\mu}} = 
	u^{\mu} \frac{\partial \ln \sqrt{|g|}}{\partial x^{\mu}} = 
	\frac{\dd \ln \sqrt{|g|}}{\dd \tau},
\end{equation}
where $ \sqrt{|g|} $ is the square root of the determinant of the metric, which equals the determinant of the vierbein $ |e| $. The geometric interpretation of $ \theta $ from the divergence of the velocity field, proceeds in GA in complete analogy, in particular by writing $ u^\mu $ as $ (1,0,0,0) $ in FRW-coordinates.

In GA, the covolume $ \sqrt{|g|} $ can be derived by considering the two relevant frames: the coordinate frame $ \{g_\mu\} $, and the orthonormal tetrad frame $ \{\g_m\} $ so that the unity pseudoscalar\footnote{The pseudoscalar is the volume element constructed from a coordinate basis vector (a top-form in differential forms language).} can be constructed with the tetrad vectors,
\begin{equation}
	I = \g_t \wedge \g_1 \wedge \g_2 \wedge \g_3, \quad |I| = 1.
\end{equation}
Because there is only one pseudoscalar element in a space, the volume element of any coordinate basis vectors $ e $ must be a scalar multiplication of $ I $, and its value will be $ |e| $.
\begin{equation}\label{eq:Vierbein_PS}
	e = g_t \wedge g_1 \wedge g_2 \wedge g_3 = |e| I, \quad |e| = |g_t \wedge g_1 \wedge g_2 \wedge g_3|.
\end{equation}
We can explicitly obtain $ |e| $ by using \Cref{eq:Tetrad_Vierbein}, obtaining $ \sqrt{|g|} = |e| $ \cite{Hamilton_Notes}, which is consistent with the view of the vierbein as the ``square root of the metric". Thus, we can write \Cref{eq:Divergence} as
\begin{equation}\label{eq:Theta}
	\theta = 
	\frac{\dd \ln \sqrt{|g|}}{\dd \tau} = 
	\frac{1}{|e|}\frac{\dd |e|}{\dd \tau} = 
	\frac{|e|'}{|e|}.
\end{equation}
paving the way to interpreting $ \theta $ as the relative variation of the volume element of the coordinate basis with respect to the proper time. In this way, one arrives at the geometric interpretation of the left-hand side of Raychaudhuri's equation as
\begin{equation}\label{eq:d_Theta}
	\frac{\dd \theta}{\dd \tau} = 
	\frac{\dd}{\dd \tau}\left(\frac{1}{|e|}\frac{\dd |e|}{\dd \tau}\right) = 
	\frac{1}{|e|}\frac{\dd^2 |e|}{\dd \tau^2} - \frac{1}{|e|^2}\left(\frac{\dd |e|}{\dd \tau}\right)^2 = 
	\frac{1}{|e|}\frac{\dd^2 |e|}{\dd \tau^2} - \theta^2 .
\end{equation}
The derivative of $ \theta $ with respect to the proper time is composed of two terms, the first one is the relative acceleration in the change of the volume element, and the second one is the square of the relative variation of the volume element. With these geometric elements at hand, one can substitute \Cref{eq:Theta,eq:d_Theta} into \Cref{eq:Ray_Simplified} and re-write Raychaudhuri's equation in a manner that fully reflects its geometric content.
\begin{equation}\label{eq:GA_Raychaudhuri_Eq_2}
	\frac{ |e| ''}{|e|} = \frac{2}{3} \left(\frac{|e|'}{|e|}\right)^2 - R(g_t) \cdot g_t+ D \cdot \left(u \cdot D u\right)
\end{equation}
with $ |e|' = \dd |e|/\dd\tau $ the derivative of the covolume with respect to the proper time.

\subsection{Friedmann's second equation as a particular case}\label{sec:friedmanns-second-equation-as-a-particular-case}
In an FRW-universe, the last term in \Cref{eq:GA_Raychaudhuri_Eq_2} vanishes due to homogeneity: In fact, the Euler-equation for the motion of the cosmological fluids is trivially fulfilled as there are no pressure gradients on spatial hypersurfaces. As a consequence, all fluid elements need to follow geodesics, which are defined through the autoparallelity condition $u^\nu\nabla_\nu u^\mu = 0$ which in GA reads as $ u \cdot D u = 0 $, producing
\begin{equation}\label{eq:Ray_Simplified}
	\frac{ |e| ''}{|e|} = \frac{2}{3} \left(\frac{|e|'}{|e|}\right)^2 - R(g_t) \cdot g_t.
\end{equation}
which requires the determination of $ |e| $ and $ R(g_t) \cdot g_t $: For obtaining $ |e| $, one can consider the conventional vierbein of a FRW-spacetime becomes
\begin{equation}
	\left[\T{e}{^m_\mu}\right]=
	\begin{pmatrix}
		1 & 0 & 0 & 0\\
		0 & a & 0 & 0\\
		0 & 0 & a & 0\\
		0 & 0 & 0 & a\\
	\end{pmatrix}.
\end{equation}
in FRW-coordinates and determinant can be derived immediately from \Cref{eq:Vierbein_PS}, giving
\begin{equation}
	|e| = a^3
\end{equation}
such that volumes increase proportionally to $a^3$ in the Hubble-expansion, as intuitively expected. Furthermore, FRW-spacetimes possess the peculiarity that proper time and coordinate time are equal, $ \tau = t $: The evolution of the volume element $|e|$ solely depends on time, and one can change the derivatives with respect to proper time time in \Cref{eq:Ray_Simplified} by derivatives with coordinate time, $ |e|' ~\rightarrow~\dot{|e|}$, resulting in
\begin{equation}
	3\frac{\ddot{a}}{a} = -R(g_t)\cdot g_t,
\end{equation}
with the curvature term, $ R(g_t) \cdot g_t $, given from \Cref{eq:Einstein_Eq} by setting $ a = g_t $, and projecting into the observer's 4-velocity $ g_t $,
\begin{equation}\label{eq:Ricci_Projection}
	R(g_t) \cdot g_t = \kappa T(g_t)  \cdot g_t + \Lambda g_t \cdot g_t + \frac{\mathcal{R}}{2} g_t \cdot g_t,
\end{equation}
where $ g_t \cdot g_t = g_{tt} = +1 $. For continuing, one requires explicit expressions for $ \mathcal{R} $ and $ T(g_t) \cdot g_t $. The latter term, $ T (g_t) \cdot g_t$, results directly from the expression of the energy-momentum tensor of a perfect fluid \Cref{eq:EM_Tensor_Perfect_Fluid}. For an FRW-universe in comoving coordinates, the fluid is at rest with respect to the observer's frame, which means $ v = g_t $. Therefore, one immediately arrives at
\begin{equation} \label{eq:EM_Projection}
	T(g_t) \cdot g_t = \rho.
\end{equation}
The first term $ \mathcal{R} $ requires the computation of the trace of Einstein's field equations, which in the language of geometric algebra is computed as
\begin{equation}
	g^\mu \cdot \left( R(g_\mu) - \frac{\mathcal{R}}{2} g_\mu = \kappa T (g_\mu) + \Lambda g_\mu\right)
	\quad\rightarrow\quad
	\mathcal{R} - 2\mathcal{R}= \kappa\: \mathrm{tr}(T) + 4 \Lambda 
\end{equation}
relating, as expected, the Ricci-scalar with the trace of the energy-momentum tensor \Cref{eq:EM_Tensor_Perfect_Fluid}. This trace in particular reads as
\begin{equation}
	\mathrm{tr}(T) = 
	g^\mu \cdot T(g_\mu) = g^\mu \left((\rho + p) g_t \cdot g_\mu g_t - p g_\mu\right) = 
	(\rho + p) - p + (-p g^i g_i) = \rho - 3 p.
\end{equation}
Combining both results implies for the Ricci scalar the well-known result
\begin{equation}\label{eq:Ricci_Scalar}
	\mathcal{R} = - \kappa Tr(T) - 4 \Lambda = -\kappa (\rho - 3 p )- 4 \Lambda.
\end{equation}
The projection $R(g_t) \cdot g_t$ can be written using \Cref{eq:EM_Projection,eq:Ricci_Scalar} and \Cref{eq:Ricci_Projection} as 
\begin{equation}\label{eq:Ricci_Projection_Final}
	R(g_t) \cdot g_t = 
	\kappa \rho + \frac{1}{2}\left(- \kappa (\rho - 3p) - 4 \Lambda\right) + \Lambda = 
	\frac{\kappa}{2} (\rho + 3 p) - \Lambda
\end{equation}
which gives an explicit expression for \Cref{eq:Ray_Simplified}, using the intermediate \Cref{eq:Theta,eq:Ricci_Projection_Final}:
\begin{equation}
	\frac{\ddot{a}}{a} = -\frac{4\pi G}{3} (\rho + 3 p) + \frac{\Lambda}{3}
\end{equation}
which is exactly Friedmann's second equation, with the substitution of $ \dot{a} $ from \Cref{eq:Friedmann1}.

\subsection{Varying the Friedmann action}\label{sec:varying-the-friedmann-action}
A third method to obtain Friedmann's equations is by performing variations on the gravitational action, which is already symmetry reduced according to the cosmological principle. Interchanging variation and symmetry reduction is permissible under certain assumptions, the most notable being a compact symmetry group, but this condition is only necessary and not sufficient \citep{fels_principle_2002, torre_symmetric_2011}. The action of a FRW-spacetime is given by the Ricci-scalar  \Cref{eq:Ricci_Scalar_FLRW} with a term called lapse function $ N $.
\begin{equation}\label{eq:Action_FLRW}
S = 
\frac{1}{8 \pi G}\int \dd^4 x\:
\left[N \left(\Lambda - \frac{3 k}{a^2}\right) + \frac{3\dot{a}^2}{N a^3} \right]
\end{equation}
so that variation with respect to $a$ and $N$ as degrees of freedom recovers the Friedmann-equations.

The purpose of the lapse function is to introduce the freedom to choose the time parameterization, and it is put in place as the $ g_{tt} $-term of the metric. At first sight, \Cref{eq:Action_FLRW} looks the same as in conventional tensor formalism. However, notice the absence of $ \sqrt{|g|} $. This is because $ \dd^4 x $ is an oriented differential, which means that we can decompose it as 
\begin{equation}
	\dd^4 x = g_0 \wedge g_1 \wedge g_2 \wedge g_3 |\dd x^0| |\dd x^1| |\dd x^2| |\dd x^3| = |e| I |\dd^4x|.
\end{equation}
\noi Where $ g_0 \wedge g_1 \wedge g_2 \wedge g_3 $ is the pseudoscalar construced from of the coordinates frame and $ |\dd^\mu x| $ are the coordinate differentials used to perform the integration. Remember from \cref{sec:Raychaudhuri-congruences} that every pseudoscalar in a space is a scalar multiplication of the unit pseudoscalar $ I $. In this case, the scale factor is the volume element encased by our coordinate basis vectors $ |e| = |g_0 \wedge g_1 \wedge g_2 \wedge g_3| $, according to \Cref{eq:Vierbein_PS}. In FRW-coordinates $ |e| = a^3 $ and in conformal coordinates $ |e| = a^4 $.

\section{Conformal flatness of FRW-spacetimes}\label{sec:conformal-flatness-of-flrw-spacetimes}
FRW-spacetimes are conformally flat, meaning that their metric $g_{\mu\nu}$ result from the Minkowski metric $\eta_{\mu\nu}$ by scaling with a position dependent, strictly positive factor $\alpha^2(x)$,
\begin{equation}\label{eq:Conformal_Flatness}
	g_{\mu\nu} = \alpha^2(x)\eta_{\mu\nu}.
\end{equation}
This implies that, due to the symmetries of the problem, the tetrads can be derived from a particular set of coordinates. This is not a general result, and usually an orthonormal frame needs to be defined locally as a transformation of the coordinate frame.

In particular, null-geodesics, characterised by a vanishing line element $\dd s^2 = 0$, are invariant under conformal transformations, and conformally flat spacetimes allow a coordinate choice with manifestly Lorentzian light cones: The FRW-line element is commonly expressed in comoving coordinates and physical (or cosmic) time, as it measures the length of the world lines of comoving observers, as
\begin{equation}\label{eq:Flat_FLRW}
	\dd s^2 = \dd t^2 -a^2 (t) \left[\dd r^2 +  r^2 \dd\Omega^2\right].
\end{equation}
Introducing conformal time $\dd\eta$ through
\begin{equation}\label{eq:Conformal_transformation_FLRW}
	\dd\eta = \frac{\dd t}{a(t)},
\end{equation}
the line element is reduced to that of Minkowski-spacetime. Here, in spherical coordinates,
\begin{equation}\label{eq:Conformally_Flat_FLRW}
	\dd s^2 = a^2 (\eta)\left[ \dd \eta^2 - \dd r^2 +  r^2 \dd\Omega^2\right]
\end{equation}
with the scale factor $a(t)$ playing the role of the conformal factor $\alpha(x)$, which only depends on time in this case. This reduction can be done in curved FRW spacetimes as well; spatial flatness is not a necessity for conformal flatness.

\subsection{Weyl curvature}
Conformal transformations leave the Weyl curvature invariant, and conformal flatness implies a vanishing Weyl curvature. These two properties have particular relevance to FRW-spacetimes, as the symmetries of the cosmological principle, spatial homogeneity and isotropy, require the Weyl tensor to vanish. Because the dynamics with the scale factor $a(t)$ is a mere conformal transformation, the Weyl-tensor remains zero in time evolution. This is just another way of expressing the fact that the Hubble expansion maintains the FRW symmetries, commonly formulated in the way that the cosmological fluids remain at rest in the comoving frame. The condition for spatial flatness, i.e. that the densities of the fluids add up to $3H(t)^2/8\pi G$, is independent of the cosmological symmetries, so it is no surprise that spatial flatness and conformal flatness are two independent concepts. 

In GA, the Weyl tensor is derived from algebraic arguments\cite{Weyl_StackExchange, Lasenby_Doran_Gull_1998}. A possible starting point is considering the following property of the Ricci tensor:
\begin{equation}
	\p_a \cdot ( R(a) \wedge b) = \p_a \cdot R(a) b - b \cdot \p_a R(a) = \mathcal{R} b - R(b),
\end{equation}
with $\p_a$ being the derivative with respect to the vector $a$, not to be confused with $a \cdot D$, which is the directional derivative in the $a$ direction. $\p_a$ is mathematically equivalent to $\nabla$, but with the argument where it acts upon made explicit \cite{Hestenes1996}.

Because the Riemann tensor is a function of $a \wedge b$ such that $\p_a \cdot R(a \wedge b) = R(b)$, one term must be $R(a) \wedge b$, and to satisfy the anti-symmetry of the argument, it must have also a term $a \wedge R(b)$. Applying $\p_a \cdot $ to this term one gets 
\begin{equation}
	\p_a \cdot (R(a) \wedge b + a \wedge R(b) = 2 R(b) + b \mathcal{R}.
\end{equation}
Furthermore, noting that
\begin{equation}
	\p_a (a \wedge b) = 4b - b = 3b, 
\end{equation}
one arrives at
\begin{equation}
	\p_a \cdot [\frac{1}{2}(R(a) \wedge b + a \wedge R(b) - \frac{1}{6}\mathcal{R} a \wedge b] = R(b) = \p_a \cdot R(a \wedge b).
\end{equation}
Therefore, it is possible to rewrite the Riemann tensor as 
\begin{equation}
	R(a \wedge b) = \frac{1}{2}(R(a) \wedge b + a \wedge R(b)) - \frac{1}{6}\mathcal{R} a \wedge b + C(a \wedge b),
\end{equation}
where $C(a \wedge b)$ is an arbitrary function satisfying $\p_a \cdot C(a \wedge b) = 0$ which is called the Weyl tensor. One can proceed by isolating this particular tensor,
\begin{equation}
	C(a \wedge b) \equiv R(a \wedge b) - \frac{1}{2}\left(R(a) \wedge b + a \wedge R(b) - \frac{1}{6} a \wedge b R\right).
\end{equation}
and it can be proven that besides being traceless, it fulfils the property
\begin{equation}
	\p_a C(a \wedge b) =  \p_a \cdot  C(a \wedge b) + \p_a \wedge C(a \wedge b) = 0.
\end{equation}
Its application to the bivector $g_\mu \wedge g_\nu $ and its decomposition into a coordinate base provides the conventional components of the Weyl tensor in tensorial formalism
\begin{equation}
	C (g_\mu \wedge g_\nu) = C_{\mu\nu} = \frac{1}{2} C_{\mu\nu \alpha \beta }\: g^\alpha \wedge g^\beta.
\end{equation}
Unfortunately, there does not seem to be a computational advantage over conventional Riemannian geometry to prove that the Weyl-tensor is zero for FRW-spacetimes, nor to show its invariance under conformal transformations.

\subsection{Conformal transformations in GA and conformal flatness}
There is a notational advantage in GA with respect to conformal transformations, though: The relative change of the coordinate basis vectors $ \{g_\mu\} $ due to curvature can be easily written as
\begin{equation}
	g_{\mu\nu} = g_\mu \cdot g_\nu,
\end{equation}
where each of the basis vectors is obtained by applying the directional derivative on the coordinate function $x$,
\begin{equation}
	g_\mu = \frac{\p x}{\p x^\mu} = \p_\mu x,
\end{equation}
such that we can obtain their reciprocal frame by applying the vector derivative $ \nabla = g^\mu \p_\mu $ to each of the inverse scalar mappings $ x^\mu = x^\mu (x) $
\begin{equation}
	g^\mu = \nabla x^\mu.
\end{equation}

Therefore, the line element is the product of two differential vectors,
\begin{equation}
	\dd s^2 = 
	g_{\mu \nu} \dd x^\mu \dd x^\nu = 
	\dd x^{(1)} \cdot \dd x^{(2)}, 
	\quad\text{with}\quad
	\dd x^{(i)} = (\dd x^\mu)^{(i)} g_\mu^{(i)}
\end{equation}
where $ \dd x^{(i)}$ is a differential vector with scalar components $ \dd x^\mu $. Conformal transformations of the metric with a conformal factor $\alpha^2(x)$ are now equivalent to transformation of the coordinate differential $\dd x$ as a GA-vector with a single power of $\alpha(x)$:
\begin{equation}
	x \mapsto x'
	\quad\text{implies}\quad 
	g'_\mu = \p_\mu x' = \alpha(x) g_\mu
\end{equation}
for any $\alpha(x) \in \mathbb{R}$.
Then, we can write the new metric as a scalar multiplication of the old one,
\begin{equation}
	g'_{\mu\nu} = g'_\mu \cdot g'_\nu = \alpha(x)^2 g_{\mu\nu}.
\end{equation}

We can also straightforwardly obtain the change in the volume element after a conformal transformation. As explained in \Cref{sec:varying-the-friedmann-action}, the volume element $ \sqrt{|g|} $ is simply the coordinate volume element and is equivalent to the determinant of the vierbein $ e $
\begin{equation}
	e =	g_0 \wedge g_1 \wedge g_2 \wedge g_3 = |g_0 \wedge g_1 \wedge g_2 \wedge g_3| I = \sqrt{|g|} I,
\end{equation}
with $ I = \g_0 \wedge \g_1 \wedge \g_2 \wedge \g_3 $ the unit pseudoscalar formed by the orthonormal basis $ \{\g_m\} $.

After a conformal transformation, the new volume element $e'$ is
\begin{equation}
	e'= g'_0 \wedge g'_1 \wedge g'_2 \wedge g'_3 = \alpha(x)^4 g_0 \wedge g_1 \wedge g_2 \wedge g_3 = \alpha(x)^4 e.
\end{equation}
 Where we can see that a conformal transformation changes the volume element as $ \sqrt{|g'|} \rightarrow \alpha(x)^4 \sqrt{|g|}$.

In the FRW universe, the volume element $e$ in FRW-coordinates was obtained in \Cref{sec:friedmanns-second-equation-as-a-particular-case}, $|e(t)| = a^3 (t) $. Under the conformal transformation \Cref{eq:Conformal_transformation_FLRW}, the new volume element will be
\begin{equation}
	|e' (\eta)| = a(\eta)^4.
\end{equation}
In both cases, we can see that the volume element only depends on the time parameter, reflecting the isotropy and homogeneity assumptions.

\section{Spacetime symmetries and conservation laws}
\label{sec:spacetime-symmetries-and-conservation-laws}
FRW spacetimes are highly symmetric due to the shift- and rotation invariance required by the cosmological principle. There are different possible interpretations of how these symmetries are maintained in the course of the time evolution, or equivalently, how the Hubble expansion is the only dynamical evolution of FRW-spacetime that is compatible with the cosmological symmetries: From a geometric point of view one could argue that the scale factor introduces a conformal scaling of the metric leaving the Weyl-curvature invariant and in fact zero, making sure that the FRW-spacetime pertains only Ricci-curvature in agreement with the cosmological principle. From the point of view of fluid mechanics, diluting the cosmological fluids with the scale factor would lead to the only admissible continuity equation that would not change the spatial uniformity of the fluids, ensured by a trivially fulfilled Euler-equation.

\subsection{Energy-momentum conservation, continuity and Euler Equations}
Conservation of energy and momentum can be expressed in terms of a continuity equation, which in GA takes on the form
\begin{equation}\label{eq:Conservation_EM_1}
	D \cdot T(a) = 0.
\end{equation}
\noi Because $ T(a) $ is a symmetric tensor, $ T(a) $ and its adjoint $ \overline{T}(a) $, are identical. By using the identity, $ T(a)\cdot b = \overline{T}(b) \cdot a = T(b) \cdot a$ one can rewrite \Cref{eq:Conservation_EM_1} as
\begin{equation}\label{eq:Energy-momentum}
	D \cdot T(a) = a \cdot T(D) = 0 \Rightarrow T (D) = 0.
\end{equation}
Specifying for the energy-momentum tensor of a perfect fluid, \cref{eq:EM_Tensor_Perfect_Fluid}, and making $D$ act left and right, we obtain
\begin{equation}
	(u \cdot \acute{D})(\acute{\rho} + \acute{p}) u + (\rho + p) u (\acute{D} \cdot \acute{u}) + (\rho + p) (u \cdot \acute{D}) \acute{u} - D p = 0
\end{equation}
Where we used the tilde to denote over which terms $D$ acts. Notice that all terms are vectors, including last term which is the gradient of $p$.

We can reduce this expression by making the following considerations: third term vanishes due to geodesic equation $(u \cdot D) u = 0$, expanding the covariant divergence in second term, with \cref{eq:Divergence}, choosing in FRW-coordinates with $|e| = a^3$, being $u = \g_t$ a 4-velocity and considering that the gradient of $p$ reduces to its time derivative due to isotropy. Then, we obtain the continuity equation for FRW
\begin{equation}
	\dot{\rho} + 3 \frac{\dot{a}}{a}\left(\rho + p\right) = 0.
\end{equation}

\subsection{Lie derivatives}
The point of whether spacetime exhibits certain symmetries is independent of the coordinate choice, which might or might not be adapted to the symmetries at hand. In either case, due to the diffeomorphism invariance, general relativity is perfectly capable to determine the geometric and dynamical properties of spacetime. It is possible, though, to find directions in which a geometric object like the metric is shift-invariant, and this invariance corresponds exactly to a vanishing Lie derivative. 

Because the Lie derivative is a pre-metric construction and Clifford bundles depend on the metric to be defined, defining Lie derivatives in general on a Clifford bundle is a subtle issue, see \cite{Hestenes1996,Rodrigues_de_Oliveira_2007,Hannes_Thesis} for details. 

The Lie derivative of a vector $ a $, in the direction of the vector $ v $, is given by 
\begin{equation}
	\mathcal{L}_v a = v \cdot \nabla a - a \cdot \nabla v,
\end{equation}
where $ \nabla = \g^\mu \p_\mu$ is the differential operator in flat spacetime and, therefore, only contains partial derivatives.

If the spacetime is torsion-free and metric-compatible, one can choose the Levi-Civita connection and write the Lie derivative in terms of the covariant derivative $ D $
\begin{equation}
	\mathcal{L}_v a = v \cdot D a - a \cdot D v.
\end{equation}
This definition can be generalised to obtain the Lie derivative of multivector fields\footnote{See \cite{Hestenes1996} for the expression for comultivector fields}, which we will need to find the Lie derivative of the metric,
\begin{equation}\label{eq:Lie_Multivector}
	\mathcal{L}_v A = v \cdot D A - \acute{v} \wedge (\acute{D} \cdot A).
\end{equation}
Because, in general, the Lie derivative does not commute with the metric, it does not follow the Leibnitz rule either with the inner or the geometric product. However, it does so with the outer product because it is a metric-independent operation.
\begin{equation}
	\mathcal{L}_v (A \wedge B) = (\mathcal{L}_v A) \wedge B + A \wedge (\mathcal{L}_v B).
\end{equation}

We can use the Lie derivative of a multivector to define its action on a tensor field $ T(a,b) $ as
\begin{equation}
	\mathcal{L}_v T(a,b) = \mathcal{\dot{L}}_v \dot{T}(a,b) = \mathcal{L}_v[T(a,b)] - T(\mathcal{L}_v a, b) - T(a, \mathcal{L}_v b)
\end{equation}
where $ \mathcal{L}_v[T(a,b)] $ means evaluating the derivative as a tensor function -- that is, deriving also the arguments. 

An interesting application of the Lie derivative is that of the metric tensor $ g(a,b) $ along with the vector $ \xi $
\begin{equation}\label{eq:Lie_Metric}
	\begin{split}
		\mathcal{L}_\xi g(a,b) & = \xi \cdot D (a \cdot b) - g(\mathcal{L}_\xi a, b) - g(a, \mathcal{L}_\xi b) \\
		& = (\xi \cdot D a) \cdot b + a \cdot (\xi \cdot D b) 
		- b \cdot (\xi \cdot D a - a \cdot D \xi) - a \cdot (\xi \cdot D b - b \cdot D \xi)\\
		& = a \cdot (b \cdot D \xi) + b \cdot (a \cdot D \xi).
	\end{split}
\end{equation}

Spacetime symmetries would now correspond to directions along which the metric does not change. Observing \Cref{eq:Lie_Metric}, we see that $ \mathcal{L}_\xi g(a,b) = 0 $ if and only if $ \xi $ satisfies
\begin{equation}\label{eq:Killing_eq}
	a \cdot (b \cdot D \xi) + b \cdot (a \cdot D \xi) = 0,
\end{equation}
which is known as Killing equation\cite{Misner1973}, and $ \xi $ is called a Killing vector. The usual form of it can be obtained replacing $ (a,b) ~\rightarrow~(g_\mu, g_\nu) $
\begin{equation}
	g_\mu \cdot (g_\nu \cdot D \xi) + g_\nu \cdot (g_\mu \cdot D \xi) = 0 ~ \Rightarrow ~(\p_\nu \xi)_\mu + ( \p_\mu \xi)_\nu = 0,
\end{equation}
As such, the Killing equation is an eigenvalue equation. This allows us to isolate the symmetries of spacetime in terms of the Killing vectors. Then, we can use that knowledge to choose a coordinate system where one of the killing vectors is a coordinate basis. In that coordinate system, the metric would not depend on that particular coordinate.

\subsection{Killing vectors}

From the GA framework, we can obtain some of the properties of Killing vectors fields immediately. First, \Cref{eq:Killing_eq} means that $ a \cdot D \xi $ is antisymmetric. Therefore, $ D \xi $ is completely determined by its protraction\cite{Hestenes1996}, resulting in the bivector $ \Omega $,
\begin{equation}\label{eq:Killing_D}
	D\xi = D \wedge \xi = 2\Omega.
\end{equation}
By recalling that bivectors are the generators of the Lorentz group, it follows that the the bivectors $\Omega$ are the generators of the symmetries associated with $\xi$.

\Cref{eq:Killing_D} automatically means that Killing vectors are divergence-free,
\begin{equation}\label{eq:Killing_Divergence}
	D \cdot \xi = 0.
\end{equation}
This reflects that Killing vector fields are the associated with of a one-parameter family of curves, the trajectories of the isometry, along which the geometry is invariant\cite{Misner1973}. If the Killing vector field would not be divergence-free, the trajectories of the isometry may converge or expand, and thus would not keep lengths invariant.

This can be seen by exploring the Lie Derivative of the pseudoscalar $ I $ along a vector field $ v $. By applying  \cref{eq:Lie_Multivector} we obtain
\begin{equation}\label{eq:Lied_Pseudoscalar}
	\mathcal{L}_v I = - I D \cdot v.
\end{equation}
Meaning that the variation in the pseudoscalar $I$ along the direction of $v$ is governed by the divergence of the vector field $v$.

From \Cref{eq:Killing_D} we see that Killing's equation is equivalent to the requirement of $ \xi $ to satisfy
\begin{equation}\label{eq:Killing_Directional}
	a \cdot D \xi = a \cdot \Omega.
\end{equation}
This can easily be checked by expanding in coordinates,
\begin{equation}
	\begin{aligned}
		\Omega &= \frac{1}{2}\left(D_\mu \xi_\nu - D_\nu \xi_\mu\right) g^\mu \wedge g^\nu\\
		g_\alpha \cdot \Omega & = \frac{1}{2}\left(D_\alpha \xi_\nu - D_\nu \xi_\alpha\right) g^\nu = 
		D_\alpha \xi_\nu g^\nu = D_\alpha \xi
	\end{aligned}
\end{equation}
in the last step, we used Killing's equation \Cref{eq:Killing_eq}.

As a note, in \cite{Sobczyk_Killing} \Cref{eq:Killing_Directional} is used to define the Killing vector fields in GA, which is a different route to the one taken here of deriving them from the requirement of vanishing Lie derivative of the metric.

\subsubsection{Killing vectors of FRW-spacetimes}

Solving Killing's equation for a given spacetime is in general a difficult task. However, we can use the known symmetries of the spacetime to try to guess its Killing vectors. For FRW this is an easy task. Due to the cosmological assumptions of isotropy and homogeneity, we know that the 3-dimensional time slices are maximally symmetric. Therefore, we can use the Killing vectors of the Euclidean space $\mathbb{E}^3$, which are easily obtained in cartesian coordinates \cite{NathalieDeruelle1382}, and transform them to spherical coordinates.

The FRW universe exhibits 6 Killing vectors, corresponding to 3 translations and 3 rotations in each time slice. Any linear combination of Killing vectors is a Killing vector, so we can express the general Killing vectors of FRW as
\begin{equation}
	\xi = \sum_{i=1}^{3}\alpha_i \xi^{(i)} + \sum_{j=1}^{3}\beta_j \xi^{(j)}
\end{equation}
where $\xi^{(i)}$ are the Killing vectors related to translations and $\xi^{(j)}$ are related to rotations.

For the flat FRW universe, the components of the general Killing vector $\xi$ are
\begin{equation}\label{eq:General_Killing_FLRW}
	\begin{aligned}
		\xi^t & =0 \\
		\xi^ r & =\alpha_3 \cos (\theta )+\sin (\theta ) (\alpha_2 \sin (\phi )+\cos (\phi ) (\alpha_1 + 2 \beta_3  r \cos (\theta ))) 	\\
		\xi^\theta & =\frac{1}{ r}\left(\cos (\theta ) (\alpha_1 \cos (\phi )+\alpha_2 \sin (\phi ))-\alpha_3 \sin (\theta )+\beta_2  r \sin (\phi )+\beta_3  r \cos (2 \theta ) \cos (\phi )\right)\\
		\xi^\phi & =\frac{1}{ r \sin \theta}\left(-\sin (\phi ) (\alpha_1+\beta_3  r \cos (\theta ))+\cos (\phi ) (\alpha_2+\beta_2  r \cos (\theta ))+\beta_1  r \sin (\theta )\right).
	\end{aligned}
\end{equation}

Notice that there is no time-like Killing vector, reflecting the fact that the FRW universe is curved in the time direction.

As an example, we will show the case for $\alpha_1 = 1, \alpha_{i\neq 1}= \beta_i=0$, which corresponds to a constant translation in the cartesian $x$-direction,
\begin{equation}
	\xi^{(1)} =  \sin (\theta ) \cos (\phi ) g_ r +\frac{\cos (\theta ) \cos (\phi )}{r}g_\theta-\frac{\csc (\theta ) \sin (\phi )}{r} g_\phi.
\end{equation}
And  we obtain the corresponding bivector by rising the indices of $\xi^{(1)}$ and using \Cref{eq:Killing_D}
\begin{equation}
	\begin{aligned}
		\Omega^{(1)} & = a \dot{a}\sin (\theta )  \cos (\phi ) g^t \wedge g^ r\\
		& + a \dot{a}  r \cos (\theta ) \cos (\phi ) g^t \wedge g^\theta\\
		& - a \dot{a}  r \sin (\theta ) \sin (\phi ) g^t \wedge g^\phi.
	\end{aligned}
\end{equation}

\subsubsection{Conserved quantities}
By Noether's theorem, we know that any continuous symmetry is associated with a conserved quantity. For an isometry of the metric, the conserved quantity associated with it is the projection of the Killing vector along a geodesic with tangent vector $ v $,
\begin{equation}
	v \cdot D (v \cdot \xi) = (v \cdot Dv) \cdot \xi + v \cdot (v \cdot D \xi) = v \cdot (v \cdot \Omega) = (v \wedge v) \cdot \Omega = 0.
\end{equation}
where $(v \cdot Dv) \cdot \xi$  vanishes due to auto-parallelity expressed by the geodesic equation, $ v \cdot D v = 0 $. The last term vanishes too because $ v \wedge v = 0 $.

The prior example provides an illustrative case. If we express $\xi^{(1)}$ in cartesian coordinates $(t,x,y,z)$, $\xi^{(1)} = g_x$. For a geodesic with tangent vector $v$
\begin{equation}\label{eq:Killing_Conservation}
	v \cdot D (v \cdot \xi) = v \cdot D (v_x) = 0.
\end{equation}
Meaning that the $v_x$ component is conserved along the geodesic. If we take into account that free-falling bodies follow geodesics, \Cref{eq:Killing_Conservation} means that the momentum in the $x$-direction is conserved.

Repeating the argument for the general Killing vector $\xi$ in \Cref{eq:General_Killing_FLRW} we find that linear and angular momentum are conserved for each time slice of the FRW universe.

Another conserved quantity along Killing vector fields was obtained in \Cref{eq:Lied_Pseudoscalar}, where we saw that by taking into account that Killing vector fields are divergence-less we obtain
\begin{equation}
	\mathcal{L}_\xi I = - I D \cdot \xi = 0.
\end{equation}

Meaning that, along Killing vector fields, the spacetime geometry is invariant and therefore the pseudoscalar is conserved.

\subsubsection{Tensor symmetries}

Another important consequence of the existence of Killing vector fields is the restriction that they impose on the degrees of freedom of the Riemann tensor. Consider the following equation relating covariant derivatives and the Riemann tensor,
\begin{equation}
	R(a \wedge b) \cdot c = b \cdot D(a \cdot D c) - a \cdot D(b \cdot D c) - (\mathcal{L}_a b) \cdot D c = c_{;ab} - c_{;ba}.
\end{equation}
Where we have introduced the common notation, $ a \cdot D c = c_{;a} $.

If choose $ c $ to be a Killing vector $ \xi $, we obtain
\begin{equation}
	R(a \wedge b) \cdot \xi = 
	b \cdot D(a \cdot D \xi) - a \cdot D(b \cdot D \xi) = 
	b \cdot D(a \cdot \Omega) - a \cdot D(b \cdot \Omega).
\end{equation}
Dotting with a general vector $ c $ and using $ R(A) \cdot B = R(B) \cdot A $ we arrive at
\begin{equation}
	R(a \wedge b) \cdot (\xi \wedge c) = (a \wedge b) \cdot R(\xi \wedge c) =  
	a \cdot (\Omega_{;b}) \cdot c - b \cdot (\Omega_{;a}) \cdot c.
\end{equation}
By taking derivatives with respect to $ a $, followed by derivatives with respect to $ b $ and then using
\begin{equation}
	D \wedge D \wedge \xi =  0 ~ \leftrightarrow ~ \Omega_{;a} + \p_b \wedge (\Omega_{;b} \cdot a)= 0,
\end{equation}
we find
\begin{equation}\label{eq:Riemann_Killing 1}
	R(c \wedge \xi) = -\p_b \wedge (\Omega_{;b} \cdot c ) = 
	-(\p_b \wedge \Omega_{;b}) \cdot c + \Omega_{;c} = 
	c \cdot D \Omega.
\end{equation}
From this relation, it is straightforward to recover the known equation relating the second directional covariant derivative of Killing vectors and the Riemann tensor
\begin{equation}
	R(\xi \wedge c) \cdot a = \xi_{;ac} = c \cdot \acute{D} (a \cdot D \acute{\xi}) .
\end{equation}

Another useful equation can be obtained by contracting \Cref{eq:Riemann_Killing 1}
\begin{equation}\label{eq:Laplacian_Killing}
	R(\xi) = D \cdot \Omega = D \Omega = D^2 \xi.
\end{equation}

The last proof that we present here is that the Lie derivative of a Killing vector is also a Killing vector,
\begin{equation}
	\xi_3 = \mathcal{L}_{\xi_1} \xi_2 = \xi_1 \cdot D \xi_2 - \xi_2 \cdot D \xi_1 = \xi_1 \cdot \Omega_2 - \xi_2 \cdot \Omega_1.
\end{equation}
Calculating now the directional derivative of $ \xi_3 $,
\begin{equation}
	\begin{split}
		a \cdot D \xi_3 & = (a \cdot \Omega_1) \cdot \Omega_2 - (a \cdot \Omega_2) + \xi_1 \cdot \Omega_{1;a} - \xi_2 \cdot \Omega_{2;a}\\
		& = a \cdot (\Omega_1 \times \Omega_2) + \xi_1 \cdot R(a \wedge \xi_2) - \xi_2 \cdot R (a \wedge \xi_1)\\
		& = a \cdot (\Omega_1 \times \Omega_2) + a \cdot 2R(\xi_1 \wedge \xi_2) = a \cdot \Omega_3,
	\end{split}
\end{equation}
one sees that $ \xi_3 $ is a Killing vector with bivector $ \Omega_3 =  (\Omega_1 \times \Omega_2) + 2R(\xi_1 \wedge \xi_2)$.

\section{Quintessence Lagrange density}\label{sec:quintessence-lagrangian-density}
The formalism of GA is foremost a tool for geometric objects with internal degrees of freedom and has limited additional power over conventional methods when dealing with scalar quantities. Nevertheless, because of the importance of scalar fields in cosmology, in particular at early times driving cosmic inflation and at late times in the context of quintessence dark energy, we revisit the fundamental constructions \citep{wetterich_cosmology_1988, peebles_cosmology_1988, ratra_cosmological_1988, peebles_cosmological_2003} from the point of view of GA.

\subsection{Quintessence equation of motion}
The field equation of a scalar field $\phi$ is given by
\begin{equation}\label{eq:Laplacian_Lagrangian}
	\square \phi = 
	D^2 \phi = 
	D(D\phi) = 
	D \cdot v = 
	\frac{1}{|e|}\frac{\p \left(|e| (D \phi)^\mu\right)}{\p x^\mu}.
\end{equation}
Where one can write $ D^2 \phi = D \cdot v $ because $ D \wedge D \phi = 0 $, which is equivalent to chosing the torsion-free condition $ D \wedge g^\mu = 0 $, and requiring integrability condition for scalar fields \citep{Hestenes_STG_with_GC}:
\begin{equation}
	D \wedge D \phi = 
	D \wedge g^\mu \p_\mu \phi = 
	g^\nu \wedge g^\mu \p_\nu \p_\mu \phi \overset{!}{=} 0
	\quad\rightarrow\quad 
	\p_\nu \p_\mu \phi = \p_\mu \p_\nu \phi
\end{equation}
For the particular case of FRW spacetimes, one has $ |e| = a^3$ and the homogeneity assumption only allows time derivatives, $\p_\mu ~ \rightarrow~ \p_t $, such that one can write \Cref{eq:Laplacian_Lagrangian} as
\begin{equation}
	D^{2} \phi = 
	\frac{1}{a^{3}} \partial_{t}\left(a^{3} \dot{\phi}\right)=\frac{1}{a^{3}}\left(3 \dot{a} a^{2} \dot{\phi}+a^{3} \ddot{\phi}\right) = 
	3 \frac{\dot{a}}{a} \dot{\phi}+\ddot{\phi},
\end{equation}
where the dot represents derivatives with respect to cosmic time, $ \p_t $.

The Euler-Lagrange equations for multivector fields $ \psi $ are \cite{Lasenby1993}
\begin{equation}\label{eq:mEL-Equations}
	\p_{\psi} \mathcal{L} - \acute{\left(\p_{D \psi} \mathcal{L}\right)} \acute{D} = 0,
\end{equation}
which, in the case of scalar fields $ \phi $, reduces to the familiar result
\begin{equation}\label{eq:scalar-mEL-Equations}
	\p_{\phi} \mathcal{L} - D\left(\p_{D \phi} \mathcal{L}\right)  = 0.
\end{equation}
Then, for a standard Lagrangian defining the dynamics of the scalar field $ \phi $  
\begin{equation}
	\mathcal{L} = \mathcal{L}(\phi, D\phi) = \frac{1}{2}\left(D \phi\right)^2 - V(\phi)
\end{equation}
with $ V(\phi) $ being a potential. Evaluation of the Euler-Lagrange \Cref{eq:scalar-mEL-Equations} yields
\begin{equation}
	\frac{\p \mathcal{L}}{\p \phi} = - \frac{V(\phi)}{\p \phi}
	\quad\text{and}\quad
	\frac{\p \mathcal{L}}{\p (D\phi)} = D \phi.
\end{equation}
Which immediately suggests that the equation of motion for $\phi$ is,
\begin{equation}
	\ddot{\phi} + 3 \frac{\dot{a}}{a} \dot{\phi} = - \frac{\dd V(\phi)}{\dd \phi}.
\end{equation}

\subsection{Slow-roll conditions}
Accelerated expansion takes place if the equation of state of the dominating cosmological fluid is sufficiently negative, $w = p / \rho < -1/3$, and for solving the flatness problem one needs this accelerated expansion to be maintained for a sufficiently long time. Both conditions can be formulated in terms of derivatives of the quintessence potential $V(\phi)$, which becomes quite apparent in the Raychaudhuri equation, which relates the acceleration in the change of the volume element $ |e| $ to the Ricci-curvature
\begin{equation}\label{eq:Accelerated_Expansion}
	\frac{\dd\theta }{\dd \tau} =  
	\frac{\dd^2\ln |e|}{\dd \tau^2} = 
	- \frac{\theta^2}{3} - R(g_t) \cdot g_t.
\end{equation}
As the Ricci-curvature is determined by the energy-momentum tensor, one can use \Cref{eq:Ray_Simplified} to obtain constraints on $ \phi $: Specifically, rewriting the energy-momentum tensor of the field $ \phi = \phi(t) $ in the language of GA yields
\begin{equation}\label{eq:EM_Scalar_Field}
	T^\phi(a) = D \phi (a \cdot D) \phi - a \mathcal{L}.
\end{equation}
From this expression, one can recover the usual tensorial components as
\begin{equation}
	T^\phi_{\mu\nu} = 
	g_\mu \cdot T^\phi (g_\nu) = (g_\mu \cdot D) \phi (g_\nu \cdot D) \phi - g_\mu \cdot g_\nu \mathcal{L}.
\end{equation}
By component-wise comparison of \Cref{eq:EM_Scalar_Field} with the energy-momentum tensor of a perfect fluid, \Cref{eq:EM_Tensor_Perfect_Fluid}, it is possible to identify the energy density and pressure associated with the scalar field $\phi$,
\begin{subequations}\label{eq:rho_p_Scalar_Field}
	\begin{align}
		&\rho = g_t \cdot T^\phi(g_t) = (\p_0 \phi)^2 - \frac{1}{2}(D\phi)^2 + V(\phi) = \frac{1}{2}\dot{\phi}^2 + V(\phi) \label{eq:rho_Scalar_Field}\\
		&  p = \frac{1}{3}\mathrm{tr}(T^\phi) = \frac{1}{3}g^i \cdot T^\phi (g_j) = \frac{1}{2}\dot{\phi}^2 - V(\phi). \label{eq:p_Scalar_Field}
	\end{align}
\end{subequations}
Where, in the particular case of a spatially homogeneous scalar field $\phi$, it can only depend on time. Combining \Cref{eq:Ricci_Projection} with \Cref{eq:rho_Scalar_Field} and \Cref{eq:Ricci_Scalar} yields an expression for the Ricci-curvature that directly depends on the scalar field $\phi$ and its derivative $\dot{\phi}$,
\begin{equation}
	R = \kappa (\dot{\phi}^2 - 4 V (\phi))- 4 \Lambda,
\end{equation}
and similarly, one can obtain the Ricci vector over $ g_t $ as
\begin{equation}
	R(g_t) \cdot g_t = 
	\kappa T(g_t)  \cdot g_t + \frac{R}{2} + \Lambda  = \kappa (\dot{\phi}^2 - V(\phi)) - \Lambda.
\end{equation}
Therefore, the evolution of the volume element as given by \Cref{eq:Accelerated_Expansion} results in 
\begin{equation}
	\frac{\dd^2\ln |e|}{\dd \tau^2} = 
	- \frac{\theta^2}{3} - \kappa (\dot{\phi}^2 - V(\phi)) + \Lambda,
	\label{eq:QE_acceleration}
\end{equation}
with accelerated expansion taking place under the condition that $ \Lambda > 0$ or that the kinetic energy $\dot{\phi}^2$ of the quintessence field is smaller than its potential energy $V(\phi)$,
\begin{equation}
	\dot{\phi}^2 \ll V(\phi),
\end{equation}
for obtaining the right sign in equation \Cref{eq:QE_acceleration}.

\section{Conclusions}

In this paper we presented the description of Friedmann-Robertson-Walker spacetimes in the language of geometric algebra. GA simplifies many calculations and allows for a clear interpretation of the physical situation. We did not aim for finding unknown physical properties of FRW-spacetimes, which would be a surprise given that they are well-known and well-investigated: Rather, our intention was to apply the formalism of GA to a simple, physically well-defined system in order to demonstrate the power of the formalism and its geometric concepts. 

This is particularly true for the Raychaudhuri-equation for the evolution of congruences, i.e. of spacetime volumes bounded by geodesics. Within GA the velocity divergence $\theta = \nabla_\mu u^\mu$ turns out to be the relative expansion of the spacetime volume element, as influenced by the presence of Ricci-curvature, while vorticity and shear are absent in FRW-spacetimes. It is possible to recast the Raychaudhuri-equation into a form highly similar to the second Friedmann's equation, with the interpretation of volume evolution as a consequence of the Hubble-expansion.

The cosmological principle imposes a very high level of symmetry onto FRW-spacetimes and renders them conformally flat. Unfortunately, there does not seem to be a computational advantage to using GA in terms of conformal invariance of the Weyl-tensor. However, the discussion showed that the square root of the determinat of the metric, $ \sqrt{|g|} $, is simply the value of the coordinate volume element within GA and not the covolume in conventional Riemannian geometry. Given the highly symmetric nature of FRW-spacetimes, the role of Lie-derivatives and Killing-vectors should be clarified: One can not define a general Lie derivative due to the necessity of having to fix the metric prior in order to define a Clifford space. In Clifford spaces, the Lie derivative can be properly defined only for Killing fields. Effects of repulsive gravity caused by scalar, self-interacting fields in a state of slow roll can easily be described in GA. In particular, in combination with Raychaudhuri's equations, the slow-roll conditions are recovered.

In summary, GA proved to be an excellent coordinate-free formalism to deal with the geometry and dynamics of FRW-spacetimes. It is concise and helps could help both students and researchers to develop better intuition for the topic.

\section*{Acknowledgements}
PBP, BMS and MdK would like to thank the Vector-Stiftung for financial support in the framework of the MINT innovation programme.

\bibliographystyle{unsrt}
\bibliography{Biblio}

\clearpage
\appendix

\section{Introduction to geometric algebra}\label{sec:introduction-to-ga}

Because a detailed introduction to GA can be found in various sources, \cite{Vector_Algebra_War_Chappell, Hestenes2003a_OerstedMedal, Doran2013}, we will only briefly introduce it.

GA is motivated by the desire to find a closed algebra on a vector space with respect to multiplication. We obtain it when looking for a product over the elements of the vector space $ u, v, w \in \mathcal{V} $ that is:

\begin{itemize}
	\item Associative, $ (uv)w = u(vw) $
	\item Left-distributive $ u (v + w) = uv + uw $
	\item Right-distributive $ (v + w)u = vu + wu $
	\item Reduces to the usual inner product $ u^2 = g(u,u) $
\end{itemize}

These properties are satisfied by the \textit{geometric product}, which we can write as the sum of the inner and outer product of vectors,
\begin{equation}\label{eq:Geometric_Product}
	uv := u \cdot v + u \wedge v.
\end{equation}

Various things are noteworthy in this expression: First, we denote the geometric product by the absence of a symbol between elements. Second, $ u \cdot v $ and $ u \wedge v $ have different dimensionality, $ 0 $ and $ 2 $ respectively. Third, the inner product and the outer products are the symmetric and antisymmetric parts of the geometric product, respectively
\begin{align}
	u \cdot v & = \frac{1}{2}(uv + vu) \label{eq:Inner_Product} \\
	u \wedge v &= \frac{1}{2}(uv - vu). \label{eq:Outer_Product}
\end{align}

\noi Fourth, the geometric product is neither commutative nor anticommutative. If the vectors are parallel ($ u \propto v $), the geometric product reduces to the inner product and it is commutative; if the vectors are orthogonal ($ u \cdot v = 0 $), the geometric product is antisymmetric. In general, it is a mixture of both.

Since $ uv $ is a mixture of two elements of different dimensionality, it must belong to a space different than $ \mathcal{V} $. This space is called $ \C{p,q}(\mathcal{V}) $, with, $ p,q $ being the signature of the space, such that $ p+q=n = \text{dim}(\mathcal{V})$. $  \C{p,q}(\mathcal{V}) $ is the exterior algebra $ \bigwedge(\mathcal{V}) $ equipped with the geometric product instead of the outer product. The difference relies on the fact that the geometric product allows rising and lowering grades with a single operation, connecting all subspaces $ \bigwedge^k(\mathcal{V}) $.

The dimensionality of $ \C{p,q} $ is $ \sum_{k=0}^{p+q=n} \binom{n}{k} = 2^n$, and its elements are called \textit{multivectors}. Each \textit{k}-vector of $ \C{p,q} $ represents an oriented geometric object of the space. For example, in the Euclidean plane, $ \mathbb{E}^2 $, with base elements $ \{\sigma_1,\sigma_2\} $, we can construct the GA $ \C{2} $, with base elements $ \{1, \sigma_1, \sigma_2, \sigma_1\wedge\sigma_2\} $. Where $ 1 $ represents points (or scalars), $ \{\sigma_1, \sigma_2\} $ represent vectors and $ \sigma_{1} \wedge \sigma_2 $ represent an oriented plane. All the k-vectors of a space can be encoded in a single element called \textit{multivector}, which in the case of $ \C{2} $ would be
\begin{equation} \label{eq:Multivector_E2}
	M = M^0 + M^1\sigma_1 + M^2 \sigma_2 + M^3 \sigma_1 \wedge \sigma_2,
\end{equation}
being $ \{M^i\}_{i=0,1,2,3} \in \mathbb{R}$ scalars.

\begin{figure}[h!]
	\centering
	\def\svgwidth{0.7\textwidth}
	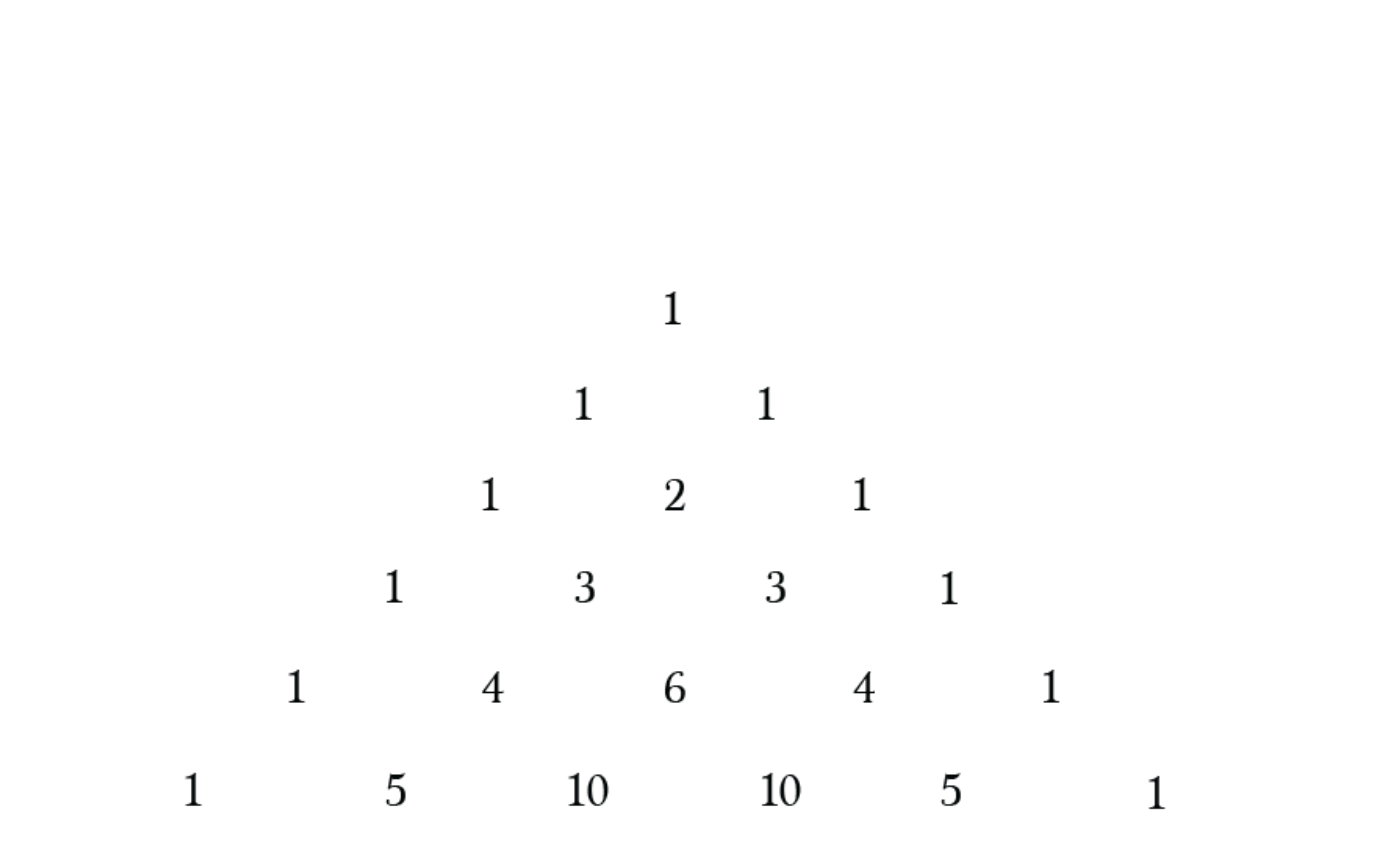
	\caption{The number of elements of a GA, $ \C{p,q} $, such that $ p + q = n $ is given by Pascal's triangle. The element with the highest dimensionality of each space is called the pseudoscalar.} 
	\label{fig:Pascal_Triangle}
\end{figure}

Similarly, in $ \C{3} $, we have $ 2^3 = 8 $ components, and a general multivector would have the form
\begin{equation}
	M = M^0 + \sum^3_{i=0} \left(M^i\sigma_i + M^i \sigma_i I\right)+ M^8 I \label{eq:Multivector_E3}
\end{equation}
with $ I = \sigma_1 \wedge \sigma_2 \wedge \sigma_3 $. The eight elements correspond to 1 scalar, 3 vectors, 3 planes(or bivectors) and 1 volume(trivector or pseudoscalar). The amount of $ k $-vectors in a Clifford space is given by Pascal's triangle, \Cref{fig:Pascal_Triangle}.

\subsection{Spacetime Algebra}\label{sec:STA}

The Minkowski spacetime in GA is called spacetime Algebra (STA)\cite{Hestenes_STA}. The basis vectors of Minkowski spacetime with signature $ (+,-,-,-) $ are
\begin{equation}
	\g_0^2 = +1, \quad \g_1^2=\g_2^2 = \g_3^2 = -1.
\end{equation}
They can be expressed compactly as
\begin{equation}\label{eq:Dirac_Algebra}
	\g_\mu \cdot \g_\nu =\frac{1}{2}(\g_\mu \g_\nu + \g_\nu \g_\mu) =\eta_{\mu\nu}.
\end{equation}
Where $ \eta_{\mu\nu} = \text{diag}(+,-,-,-)$ is the Minkowski metric. The choice of signature in Geometric Algebra is a bit more nuanced than in conventional tensor calculus, because one of the remarkable features of GA is the nicely nested structure of subalgebras that emerge naturally if we choose the $ \eta_{\mu\nu} = \text{diag}(+,-,-,-)$ signature. We call this structure the Clifford ladder and it's shown in \Cref{tab:Clifford_Ladder} \cite{Dressel2015}.

If we want to keep a similar structure with the opposite signature, $ \eta_{\mu\nu} = \text{diag}(-,+,+,+)$, we are forced to change the isomorphism between the $\C{3,1}^+$ and $\C{3}$ \cite{wu_signature_2022}. 

\begin{table}[h]
	\centering
	\begin{tabular}{c | c }
		$ Cl_{1,3} $ & spacetime Algebra\\
		$ Cl_{3,0} $ & Relative 3-space ($\mathbb{E}^3$)\\ 
		$ Cl_{0,2} $ & Quaternions \\
		$ Cl_{0,1} $ & Complex numbers \\
		$ Cl_{0,0} $ & Real numbers \\
	\end{tabular}
	\caption{Clifford Ladder. Each algebra is isomorphic to a subgroup of the higher algebra. This only stands if the signature of the spacetime is $ (+,-,-,-) $.}
	\label{tab:Clifford_Ladder}
\end{table}

The elements $ \{\g_\mu\} $ can be combined with the geometric product to construct the basis elements of  $\mathcal{C}l_{1,3}$, see \Cref{tab:STA_Basis}. As usual, the reciprocal basis, $ \{\g^\mu\} $, is defined by
\begin{equation}\label{eq:STA_Reciprocal_Bas}
	\g_\mu\g^\nu = \Tn{\delta}{^\nu_\mu}.
\end{equation}

\begin{table}[h!]
	\begin{tabular}{l | c | c | c | c | c | c }
		Scalars & 1 &  & & & \\ 
		4 vectors & $ \gamma_0 $ & $ \gamma_1 $ & $ \gamma_2 $ & $ \gamma_3 $ & & \\
		6 bivectors & $ \gamma_{10} $ & $ \gamma_{20} $ &$  \gamma_{30} $ & $ \gamma_{23} $ & $ \gamma_{31} $ & $ \gamma_{12} $\\
		4 trivectors & $ \gamma_{123} $ & $ \gamma_{230} $ & $ \gamma_{310} $ & $ \gamma_{120} $ &  \\ 
		1 Pseudoscalar &$ \gamma_{0123} $ & & & & \\
	\end{tabular}
	\caption{Basis elements of $\mathcal{C}l_{1,3}$. Notice how the scalar and the pseudo scalar, the vectors and the trivectors and the first three bivectors and the last three bivectors, are dual of each other.}
	\label{tab:STA_Basis}
\end{table}

\subsection{Rotations and boosts}\label{sec:rotations-and-boosts}

Given the aforementioned properties of the geometric product, it is easy to show that in a Euclidean space with basis vectors $ \{\sigma_i\} $, any bivector formed by the product of two orthonormal vectors $ \{\sigma_i,\sigma_j\} $ ($ \sigma_i \cdot \sigma_j = 0$, $ |\sigma_i| = 1 = |\sigma_j| $) has the algebraic properties of the imaginary unit $ i $,
\begin{equation}
	(\sigma_i \sigma_j)^2 = (\sigma_i \sigma_j)(\sigma_i \sigma_j) = \sigma_i \sigma_j\sigma_i \sigma_j = -\sigma_i \sigma_j \sigma_j \sigma_1 = - \sigma_i (\sigma_j \sigma_j) \sigma_i = -1.
\end{equation}

The geometric interpretation of bivectors and their algebraic properties makes them a handy set of generators of rotations. We can write the rotation of a multivector $ M $ in a plane and orientation described by the bivector $ \sigma_i \sigma_j $, as
\begin{equation} \label{eq:Rotation}
	M' = RM\tilde{R} = e^{\frac{\alpha}{2} (\sigma_i \sigma_j)} M e^{-\frac{\alpha}{2} (\sigma_i \sigma_j)},
\end{equation}
where $ \alpha $ is the angle of rotation. $ R $ is called a rotor and can be written as
\begin{equation}\label{eq:Rotor}
	R(\sigma_i \sigma_j) = e^{\frac{\alpha}{2} (\sigma_i \sigma_j)} = \sum_{n= 0}^{\infty} \frac{(\alpha \sigma_i \sigma_j)^n}{n!} = \cos\alpha + (\sigma_i \sigma_j)\sin\alpha.
\end{equation}

In three dimensions we have three possible bivectors corresponding to the three planes of rotation. These three bivectors can be identified with $ i = \sigma_x \wedge \sigma_y $, $ j = - \sigma_y \wedge \sigma_z $, $ k = \sigma_z \wedge \sigma_x $, and full-fill $ ijk = -1 $\footnote{Notice the minus sign in the definition of $ j $. Hamilton produced a left-handed set, which has been the origin of a lot of confusion.}. If we include the scalar element, we see that the even algebra of $ \C{3} $, called $ \C{3}^+ $, is isomorphic to quaternions.

We can use the even subalgebra of any Clifford space to generalize complex numbers to spaces of any dimension and signature. This has been done and allows us to extend complex calculus theory to manifolds of arbitrary signature\cite{Hestenes_GC}. The possibility of formulating physics without the need for complex numbers and endowing them with a geometrical meaning has been one major feature of GA, which have profound consequences when describing Quantum Mechanics with GA \cite{Doran2005_Electron_Physics,Hestenes2003b,Hestenes_Mysteries_Dirac}.

If we now return to Minkowski spacetime with basis vectors $ \{\g_\mu\} $, satisfying \Cref{eq:Dirac_Algebra}, we can see that the bivector basis of $ \C{1,3} $ is composed of 6 bivectors, \Cref{tab:STA_Basis}: 3 space-like bivectors, $ \{\g_{ij}\} $ which square to $ -1 $, and 3 spacetime bivectors, $ \{\g_{0i}\} $, which squaring to $ +1 $. Because any spacetime bivector squares to $ +1 $, when we use it to generate a rotation with \Cref{eq:Rotor}, we obtain hyperbolic functions instead of the trigonometric ones
\begin{equation}\label{eq:Rotor_spacetime}
	R(\g_i \g_0) = e^{\frac{\alpha}{2} (\g_i \g_0)} = \sum_{n= 0}^{\infty} \frac{(\alpha \g_i \g_0)^n}{n!} = \cosh\alpha + (\g_i \g_0)\sinh\alpha.
\end{equation}
We can use \Cref{eq:Rotation,eq:Rotor_spacetime} to perform boosts. E.g., a boost of the $ \g_1 $ axis with velocity $ \vec{v} \in \mathbb{R}^3 $, in the direction of $ \g_1 $ would be expressed as
\begin{equation}
	\g_1' = e^{\g_{10} \alpha/2} \g_1 e^{-\g_{10} \alpha/2} = \cosh\alpha ~ \g_1 + \sinh \alpha ~\g_0,
\end{equation}
where $ \alpha = \tanh^{-1}(|\vec{v}|/c) $.

This means that the bivector basis of a Clifford space is a representation of the Lorentz group of the underlying vector space. In the case of Minkowski $ SO (1,3) $, consistent of 3 rotations and 3 boosts. The exponential form of \cref{eq:Rotor_spacetime} makes the manipulation and composition of transformations particularly easy in comparison to the usual $ 4 \times 4 $ matrix formalism.

\subsection{Vector derivative}

In GA the fundamental derivative operator is the \textit{vector derivative}. It is constructed in analogy to the decomposition of a vector in coordinates $ v = v^\mu\g_\mu $. The vector derivative, $ \nabla $, can be defined using the reciprocal basis $ \g^\mu $, \Cref{eq:STA_Reciprocal_Bas}, as
\begin{equation}\label{eq:GA_Derivative}
	\nabla = \sum_\mu \g^\nu \overleftrightarrow{\p}_\mu,
\end{equation}
where $ \overleftrightarrow{\p}_\mu = \overleftarrow{\p}_\mu + \overrightarrow{\p}_\mu $ is the usual derivative $ \p_\mu = \p /\p x^\mu $ along the coordinate $ x^\mu = \g^\mu \cdot x $ but acting bi-sidedly.

One of the most important features of the vector operator is that we can use it algebraically as a vector and, employing the geometric product, we obtain all the conventional differential operators. 

Acting on a scalar field $ \phi = \phi(x) $, the vector derivative produces a gradient:
\begin{equation}
	\nabla \phi (x) = \g^\nu \p_\mu \phi(x).
\end{equation}

Acting on a vector field $ v = v(x) = v^\mu \g_\mu $ produces the rest of differential operators: 

\begin{itemize}
	\item Geometric product: $ \nabla v(x) = \nabla \cdot v(x) + \nabla \wedge v $
	\item Divergence: $  \nabla \cdot v = \p_\mu v^\mu$
	\item Curl:  $ \nabla \wedge v = \frac{1}{2}\sum_{\mu, \nu=0}^{3} (\p_\mu v^\nu - \p_\nu v^\mu) \g_\mu \wedge \g^\nu$
	\item Laplacian: $ \nabla^2 v = \nabla (\nabla v) = \sum_{\mu=0}^{3} \p_\mu^2 v$
\end{itemize}

Notably, acting on a bivector field $ \mathbf{F} = \mathbf{F}(x) = \mathbf{F}^{\mu \nu} \g_\mu \wedge \g_\nu$, it produces Maxwell's equations\cite{Dressel2015}.
\begin{equation}
	\begin{aligned}
		\nabla \mathbf{F}(x) &=\nabla \cdot \mathbf{F}(x)+\nabla \wedge \mathbf{F}(x) \\ &=\gamma_0\left(\partial_0+\vec{\nabla}\right)(\vec{E}+\vec{B} I) \\ &=\gamma_0\left[\vec{\nabla} \cdot \vec{E}+\partial_0 \vec{E}-\vec{\nabla} \times \vec{B}\right]+\gamma_0\left[\vec{\nabla} \cdot \vec{B}+\partial_0 \vec{B}+\vec{\nabla} \times \vec{E}\right] I.
	\end{aligned}
\end{equation}

Remarkably, the associativity property of the geometric product allows us to define the Laplacian operator not only for scalar fields but for any multivector field. The possibility of obtaining all differential operators from a primordial one is unique to Geometric Calculus. For more details we recommend \cite{Hestenes_GC,Dressel2015,Hestenes_STA}.

\end{document}

%% file: Basis_Transformation.pdf_tex
\begingroup%
  \makeatletter%
  \providecommand\color[2][]{%
    \errmessage{(Inkscape) Color is used for the text in Inkscape, but the package 'color.sty' is not loaded}%
    \renewcommand\color[2][]{}%
  }%
  \providecommand\transparent[1]{%
    \errmessage{(Inkscape) Transparency is used (non-zero) for the text in Inkscape, but the package 'transparent.sty' is not loaded}%
    \renewcommand\transparent[1]{}%
  }%
  \providecommand\rotatebox[2]{#2}%
  \newcommand*\fsize{\dimexpr\f@size pt\relax}%
  \newcommand*\lineheight[1]{\fontsize{\fsize}{#1\fsize}\selectfont}%
  \ifx\svgwidth\undefined%
    \setlength{\unitlength}{240.47018021bp}%
    \ifx\svgscale\undefined%
      \relax%
    \else%
      \setlength{\unitlength}{\unitlength * \real{\svgscale}}%
    \fi%
  \else%
    \setlength{\unitlength}{\svgwidth}%
  \fi%
  \global\let\svgwidth\undefined%
  \global\let\svgscale\undefined%
  \makeatother%
  \begin{picture}(1,0.65669698)%
    \lineheight{1}%
    \setlength\tabcolsep{0pt}%
    \put(0,0){\includegraphics[width=\unitlength,page=1]{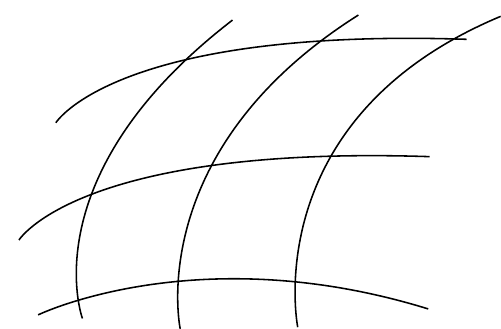}}%
    \put(0.66924193,0.62510205){\makebox(0,0)[lt]{\lineheight{1.25}\smash{\begin{tabular}[t]{l}$x^2$\end{tabular}}}}%
    \put(-0.00345189,0.20111447){\makebox(0,0)[lt]{\lineheight{1.25}\smash{\begin{tabular}[t]{l}$x^1$\end{tabular}}}}%
    \put(0,0){\includegraphics[width=\unitlength,page=2]{Basis_Transformation.pdf}}%
    \put(0.2613757,0.35276122){\makebox(0,0)[lt]{\lineheight{1.25}\smash{\begin{tabular}[t]{l}$T_p\mathcal{M}$\end{tabular}}}}%
    \put(0.42046653,0.28389673){\makebox(0,0)[lt]{\lineheight{1.25}\smash{\begin{tabular}[t]{l}$p$\end{tabular}}}}%
    \put(0,0){\includegraphics[width=\unitlength,page=3]{Basis_Transformation.pdf}}%
    \put(0.43715689,0.49230779){\makebox(0,0)[lt]{\lineheight{1.25}\smash{\begin{tabular}[t]{l}$g_2$\end{tabular}}}}%
    \put(0.57868076,0.35856597){\makebox(0,0)[lt]{\lineheight{1.25}\smash{\begin{tabular}[t]{l}$g_1$\end{tabular}}}}%
    \put(0.54860541,0.28486605){\makebox(0,0)[lt]{\lineheight{1.25}\smash{\begin{tabular}[t]{l}$\gamma_1$\end{tabular}}}}%
    \put(0.35014959,0.44891588){\makebox(0,0)[lt]{\lineheight{1.25}\smash{\begin{tabular}[t]{l}$\gamma_2$\end{tabular}}}}%
  \end{picture}%
\endgroup%

%% file: Connection_Illustration.pdf_tex
\begingroup%
  \makeatletter%
  \providecommand\color[2][]{%
    \errmessage{(Inkscape) Color is used for the text in Inkscape, but the package 'color.sty' is not loaded}%
    \renewcommand\color[2][]{}%
  }%
  \providecommand\transparent[1]{%
    \errmessage{(Inkscape) Transparency is used (non-zero) for the text in Inkscape, but the package 'transparent.sty' is not loaded}%
    \renewcommand\transparent[1]{}%
  }%
  \providecommand\rotatebox[2]{#2}%
  \newcommand*\fsize{\dimexpr\f@size pt\relax}%
  \newcommand*\lineheight[1]{\fontsize{\fsize}{#1\fsize}\selectfont}%
  \ifx\svgwidth\undefined%
    \setlength{\unitlength}{309.84233766bp}%
    \ifx\svgscale\undefined%
      \relax%
    \else%
      \setlength{\unitlength}{\unitlength * \real{\svgscale}}%
    \fi%
  \else%
    \setlength{\unitlength}{\svgwidth}%
  \fi%
  \global\let\svgwidth\undefined%
  \global\let\svgscale\undefined%
  \makeatother%
  \begin{picture}(1,0.33202563)%
    \lineheight{1}%
    \setlength\tabcolsep{0pt}%
    \put(0,0){\includegraphics[width=\unitlength,page=1]{Connection_Illustration.pdf}}%
    \put(0.23515057,0.06287457){\makebox(0,0)[lt]{\lineheight{1.25}\smash{\begin{tabular}[t]{l}$\gamma_\chi(p)$\end{tabular}}}}%
    \put(0.62282719,0.20905215){\makebox(0,0)[lt]{\lineheight{1.25}\smash{\begin{tabular}[t]{l}$\gamma'_\chi(p)$\end{tabular}}}}%
    \put(0.59285704,0.10390329){\makebox(0,0)[lt]{\lineheight{1.25}\smash{\begin{tabular}[t]{l}$\gamma_\chi(q)$\end{tabular}}}}%
    \put(0.13580947,0.23169097){\makebox(0,0)[lt]{\lineheight{1.25}\smash{\begin{tabular}[t]{l}$\gamma_t(p)$\end{tabular}}}}%
    \put(0.52478281,0.27364733){\makebox(0,0)[lt]{\lineheight{1.25}\smash{\begin{tabular}[t]{l}$\gamma'_t(p)$\end{tabular}}}}%
    \put(0.40939586,0.30750464){\makebox(0,0)[lt]{\lineheight{1.25}\smash{\begin{tabular}[t]{l}$\gamma_t(q)$\end{tabular}}}}%
    \put(0.00466673,0.00474346){\makebox(0,0)[lt]{\lineheight{1.25}\smash{\begin{tabular}[t]{l}$\chi$\end{tabular}}}}%
    \put(0,0){\includegraphics[width=\unitlength,page=2]{Connection_Illustration.pdf}}%
    \put(0.63106631,0.15890127){\makebox(0,0)[lt]{\lineheight{1.25}\smash{\begin{tabular}[t]{l}$\alpha = \dot{a}\chi$\end{tabular}}}}%
    \put(0.09648872,0.07626715){\makebox(0,0)[lt]{\lineheight{1.25}\smash{\begin{tabular}[t]{l}$p$\end{tabular}}}}%
    \put(0.46463946,0.11958471){\makebox(0,0)[lt]{\lineheight{1.25}\smash{\begin{tabular}[t]{l}$q$\end{tabular}}}}%
  \end{picture}%
\endgroup%

%% file: Curvature_Schema.pdf_tex
\begingroup%
  \makeatletter%
  \providecommand\color[2][]{%
    \errmessage{(Inkscape) Color is used for the text in Inkscape, but the package 'color.sty' is not loaded}%
    \renewcommand\color[2][]{}%
  }%
  \providecommand\transparent[1]{%
    \errmessage{(Inkscape) Transparency is used (non-zero) for the text in Inkscape, but the package 'transparent.sty' is not loaded}%
    \renewcommand\transparent[1]{}%
  }%
  \providecommand\rotatebox[2]{#2}%
  \newcommand*\fsize{\dimexpr\f@size pt\relax}%
  \newcommand*\lineheight[1]{\fontsize{\fsize}{#1\fsize}\selectfont}%
  \ifx\svgwidth\undefined%
    \setlength{\unitlength}{401.3243945bp}%
    \ifx\svgscale\undefined%
      \relax%
    \else%
      \setlength{\unitlength}{\unitlength * \real{\svgscale}}%
    \fi%
  \else%
    \setlength{\unitlength}{\svgwidth}%
  \fi%
  \global\let\svgwidth\undefined%
  \global\let\svgscale\undefined%
  \makeatother%
  \begin{picture}(1,0.50597064)%
    \lineheight{1}%
    \setlength\tabcolsep{0pt}%
    \put(0,0){\includegraphics[width=\unitlength,page=1]{Curvature_Schema.pdf}}%
    \put(0.194325,0.16311401){\color[rgb]{0,0,0}\makebox(0,0)[lt]{\lineheight{1.25}\smash{\begin{tabular}[t]{l}$A=a \wedge b$\end{tabular}}}}%
    \put(0.75726406,0.46411773){\color[rgb]{0,0,0}\makebox(0,0)[lt]{\lineheight{1.25}\smash{\begin{tabular}[t]{l}$R(a \wedge b)$\end{tabular}}}}%
    \put(0.2218504,0.00328911){\color[rgb]{0,0,0}\makebox(0,0)[lt]{\lineheight{1.25}\smash{\begin{tabular}[t]{l}$a$\end{tabular}}}}%
    \put(0.55037951,0.16755362){\color[rgb]{0,0,0}\makebox(0,0)[lt]{\lineheight{1.25}\smash{\begin{tabular}[t]{l}$b$\end{tabular}}}}%
    \put(-0.00101663,0.1613382){\color[rgb]{0,0,0}\makebox(0,0)[lt]{\lineheight{1.25}\smash{\begin{tabular}[t]{l}$b$\end{tabular}}}}%
    \put(0.27512538,0.31808065){\color[rgb]{0,0,0}\makebox(0,0)[lt]{\lineheight{1.25}\smash{\begin{tabular}[t]{l}$a$\end{tabular}}}}%
    \put(0.14282583,0.20129444){\color[rgb]{0,0,0}\makebox(0,0)[lt]{\lineheight{1.25}\smash{\begin{tabular}[t]{l}$v$\end{tabular}}}}%
    \put(0.50523989,0.41398062){\color[rgb]{0,0,0}\makebox(0,0)[lt]{\lineheight{1.25}\smash{\begin{tabular}[t]{l}$v_{ba}$\end{tabular}}}}%
    \put(0.630292,0.30873237){\color[rgb]{0,0,0}\makebox(0,0)[lt]{\lineheight{1.25}\smash{\begin{tabular}[t]{l}$v_{ab}$\end{tabular}}}}%
    \put(0.70310116,0.32382691){\color[rgb]{0,0,0}\makebox(0,0)[lt]{\begin{minipage}{0.17225582\unitlength}\raggedright \end{minipage}}}%
  \end{picture}%
\endgroup%

%% file: Pascal_Triangle.pdf_tex
\begingroup%
  \makeatletter%
  \providecommand\color[2][]{%
    \errmessage{(Inkscape) Color is used for the text in Inkscape, but the package 'color.sty' is not loaded}%
    \renewcommand\color[2][]{}%
  }%
  \providecommand\transparent[1]{%
    \errmessage{(Inkscape) Transparency is used (non-zero) for the text in Inkscape, but the package 'transparent.sty' is not loaded}%
    \renewcommand\transparent[1]{}%
  }%
  \providecommand\rotatebox[2]{#2}%
  \newcommand*\fsize{\dimexpr\f@size pt\relax}%
  \newcommand*\lineheight[1]{\fontsize{\fsize}{#1\fsize}\selectfont}%
  \ifx\svgwidth\undefined%
    \setlength{\unitlength}{659.95063913bp}%
    \ifx\svgscale\undefined%
      \relax%
    \else%
      \setlength{\unitlength}{\unitlength * \real{\svgscale}}%
    \fi%
  \else%
    \setlength{\unitlength}{\svgwidth}%
  \fi%
  \global\let\svgwidth\undefined%
  \global\let\svgscale\undefined%
  \makeatother%
  \begin{picture}(1,0.61231959)%
    \lineheight{1}%
    \setlength\tabcolsep{0pt}%
    \put(0,0){\includegraphics[width=\unitlength,page=1]{Pascal_Triangle.pdf}}%
    \put(0.53823856,0.40411487){\color[rgb]{0,0,0}\rotatebox{46.337034}{\makebox(0,0)[lt]{\lineheight{1.25}\smash{\begin{tabular}[t]{l}Scalar\end{tabular}}}}}%
    \put(0.61521945,0.33942887){\color[rgb]{0,0,0}\rotatebox{45}{\makebox(0,0)[lt]{\lineheight{1.25}\smash{\begin{tabular}[t]{l}Vector\end{tabular}}}}}%
    \put(0.67851657,0.27164121){\color[rgb]{0,0,0}\rotatebox{45}{\makebox(0,0)[lt]{\lineheight{1.25}\smash{\begin{tabular}[t]{l}Bivector\end{tabular}}}}}%
    \put(0.74816208,0.20782969){\color[rgb]{0,0,0}\rotatebox{45}{\makebox(0,0)[lt]{\lineheight{1.25}\smash{\begin{tabular}[t]{l}Trivector\end{tabular}}}}}%
    \put(0.80425612,0.13234446){\color[rgb]{0,0,0}\rotatebox{45}{\makebox(0,0)[lt]{\lineheight{1.25}\smash{\begin{tabular}[t]{l}4-vector\end{tabular}}}}}%
    \put(0.88098739,0.07184705){\color[rgb]{0,0,0}\rotatebox{45}{\makebox(0,0)[lt]{\lineheight{1.25}\smash{\begin{tabular}[t]{l}5-vector\end{tabular}}}}}%
    \put(0.2756062,0.59281912){\color[rgb]{0,0,0}\rotatebox{-45}{\makebox(0,0)[lt]{\lineheight{1.25}\smash{\begin{tabular}[t]{l}Pseudoscalar\end{tabular}}}}}%
    \put(0,0){\includegraphics[width=\unitlength,page=2]{Pascal_Triangle.pdf}}%
    \put(0.02141761,0.3737238){\color[rgb]{0,0,0}\makebox(0,0)[lt]{\lineheight{1.25}\smash{\begin{tabular}[t]{l}$\C{0}$\end{tabular}}}}%
    \put(0.01619961,0.1716657){\color[rgb]{0,0,0}\makebox(0,0)[lt]{\lineheight{1.25}\smash{\begin{tabular}[t]{l}$\C{3}$\end{tabular}}}}%
    \put(0.01722118,0.23908927){\color[rgb]{0,0,0}\makebox(0,0)[lt]{\lineheight{1.25}\smash{\begin{tabular}[t]{l}$\C{2}$\end{tabular}}}}%
    \put(0.02028589,0.3054914){\color[rgb]{0,0,0}\makebox(0,0)[lt]{\lineheight{1.25}\smash{\begin{tabular}[t]{l}$\C{1}$\end{tabular}}}}%
    \put(0.01619962,0.02966747){\color[rgb]{0,0,0}\makebox(0,0)[lt]{\lineheight{1.25}\smash{\begin{tabular}[t]{l}$\C{5}$\end{tabular}}}}%
    \put(0.01722119,0.10219891){\color[rgb]{0,0,0}\makebox(0,0)[lt]{\lineheight{1.25}\smash{\begin{tabular}[t]{l}$\C{4}$\end{tabular}}}}%
  \end{picture}%
\endgroup%